\input harvmac.tex
\input epsf

\def\figin{\epsfcheck\figin}\def\figins{\epsfcheck\figins}
\def\epsfcheck{\ifx\epsfbox\UnDeFiNeD
\message{(NO epsf.tex, FIGURES WILL BE IGNORED)}
\gdef\figin##1{\vskip2in}\gdef\figins##1{\hskip.5in}% blank space instead
\else\message{(FIGURES WILL BE INCLUDED)}%
\gdef\figin##1{##1}\gdef\figins##1{##1}\fi}
\def\DefWarn#1{}
\def\figinsert{\goodbreak\midinsert}
\def\ifig#1#2#3{\DefWarn#1\xdef#1{fig.~\the\figno}
\writedef{#1\leftbracket fig.\noexpand~\the\figno}%
\figinsert\figin{\centerline{#3}}\medskip\centerline{\vbox{\baselineskip12pt
\advance\hsize by -1truein\noindent\footnotefont{\bf
Fig.~\the\figno:} #2}}
\bigskip\endinsert\global\advance\figno by1}
%%% TO PUT FIGURES INSERT:
%\ifig\fivelegs{ Write caption
%  } {\epsfxsize2.5in\epsfbox{fivelegs.eps}}

\def\half{{1\over 2 }}

%%%%%%%%%%%%%%% References

%\KlebanovWS
\lref\KlebanovWS{
  I.~R.~Klebanov, D.~Kutasov and A.~Murugan,
  %``Entanglement as a probe of confinement,''
Nucl.\ Phys.\ B {\bf 796}, 274 (2008).
[arXiv:0709.2140 [hep-th]].
%%CITATION = arXiv:0709.2140%%
}

%\KlebanovYF
\lref\KlebanovYF{
  I.~R.~Klebanov, T.~Nishioka, S.~S.~Pufu and B.~R.~Safdi,
  %``On Shape Dependence and RG Flow of Entanglement Entropy,''
JHEP {\bf 1207}, 001 (2012).
[arXiv:1204.4160 [hep-th]].
%%CITATION = arXiv:1204.4160%%
}
%\LiBT
\lref\LiBT{
  M.~Li and Y.~Pang,
  %``Holographic de Sitter Universe,''
JHEP {\bf 1107}, 053 (2011).
[arXiv:1105.0038 [hep-th]].
%%CITATION = arXiv:1105.0038%%
}

%\BarbonTA
\lref\BarbonTA{
  J.~L.~F.~Barbon and E.~Rabinovici,
  %``AdS Crunches, CFT Falls And Cosmological Complementarity,''
JHEP {\bf 1104}, 044 (2011).
[arXiv:1102.3015 [hep-th]].
%%CITATION = arXiv:1102.3015%%
}

%\ColemanAW
\lref\CdeL{
  S.~R.~Coleman and F.~De Luccia,
  %``Gravitational Effects On And Of Vacuum Decay,''
  Phys.\ Rev.\  D {\bf 21}, 3305 (1980).
  %%CITATION = PHRVA,D21,3305;%%
}

%\AbbottKR
\lref\AbbottKR{
  L.~F.~Abbott and S.~R.~Coleman,
  %``The Collapse Of An Anti-De Sitter Bubble,''
  Nucl.\ Phys.\  B {\bf 259}, 170 (1985).
  %%CITATION = NUPHA,B259,170;%%
}

 % Talks about dS field theories and their AdS duals.
%\BuchelWF
\lref\BuchelWF{
  A.~Buchel,
  %``Gauge / gravity correspondence in accelerating universe,''
  Phys.\ Rev.\  D {\bf 65}, 125015 (2002)
  [arXiv:hep-th/0203041].
  %%CITATION = PHRVA,D65,125015;%%
}
 \lref\BuchelWF{
  A.~Buchel,
  %``Gauge / gravity correspondence in accelerating universe,''
  Phys.\ Rev.\  D {\bf 65}, 125015 (2002)
  [arXiv:hep-th/0203041].
  %%CITATION = PHRVA,D65,125015;%%
}
%\AharonyCX
\lref\AharonyCX{
  O.~Aharony, M.~Fabinger, G.~T.~Horowitz and E.~Silverstein,
  %``Clean time-dependent string backgrounds from bubble baths,''
  JHEP {\bf 0207}, 007 (2002)
  [arXiv:hep-th/0204158].
  %%CITATION = JHEPA,0207,007;%%
}
%\BalasubramanianAM
\lref\BalasubramanianAM{
  V.~Balasubramanian and S.~F.~Ross,
  %``The dual of nothing,''
  Phys.\ Rev.\  D {\bf 66}, 086002 (2002)
  [arXiv:hep-th/0205290].
  %%CITATION = PHRVA,D66,086002;%%
}
%\RossCB
\lref\RossCB{
  S.~F.~Ross and G.~Titchener,
  %``Time-dependent spacetimes in AdS/CFT: Bubble and black hole,''
  JHEP {\bf 0502}, 021 (2005)
  [arXiv:hep-th/0411128].
  %%CITATION = JHEPA,0502,021;%%
}
%\CaiMR
\lref\CaiMR{
  R.~G.~Cai,
  %``Constant curvature black hole and dual field theory,''
  Phys.\ Lett.\  B {\bf 544}, 176 (2002)
  [arXiv:hep-th/0206223].
  %%CITATION = PHLTA,B544,176;%%
}
%\BalasubramanianBG
\lref\BalasubramanianBG{
  V.~Balasubramanian, K.~Larjo and J.~Simon,
  %``Much ado about nothing,''
  Class.\ Quant.\ Grav.\  {\bf 22}, 4149 (2005)
  [arXiv:hep-th/0502111].
  %%CITATION = CQGRD,22,4149;%%
}
%\HeJI
\lref\HeJI{
  J.~He and M.~Rozali,
  %``On Bubbles of Nothing in AdS/CFT,''
  JHEP {\bf 0709}, 089 (2007)
  [arXiv:hep-th/0703220].
  %%CITATION = JHEPA,0709,089;%%
}
%\HutasoitXY
\lref\HutasoitXY{
  J.~A.~Hutasoit, S.~P.~Kumar and J.~Rafferty,
  %``Real time response on dS_3: the Topological AdS Black Hole and the
  %Bubble,''
  JHEP {\bf 0904}, 063 (2009)
  [arXiv:0902.1658 [hep-th]].
  %%CITATION = JHEPA,0904,063;%%
}
%\MarolfTG
\lref\MarolfTG{
  D.~Marolf, M.~Rangamani and M.~Van Raamsdonk,
  %``Holographic models of de Sitter QFTs,''
  arXiv:1007.3996 [hep-th].
  %%CITATION = ARXIV:1007.3996;%%
}

% AdS/CFT  on dS with mass deformation
%\BuchelKJ
\lref\BuchelKJ{
  A.~Buchel, P.~Langfelder and J.~Walcher,
  %``On time-dependent backgrounds in supergravity and string theory,''
  Phys.\ Rev.\  D {\bf 67}, 024011 (2003)
  [arXiv:hep-th/0207214].
  %%CITATION = PHRVA,D67,024011;%%
}
%\HirayamaJN
\lref\HirayamaJN{
  T.~Hirayama,
  %``A holographic dual of CFT with flavor on de Sitter space,''
  JHEP {\bf 0606}, 013 (2006)
  [arXiv:hep-th/0602258].
  %%CITATION = JHEPA,0606,013;%%
}
%\WittenGJ
\lref\WittenKK{
  E.~Witten,
  %``Instability Of The Kaluza-Klein Vacuum,''
  Nucl.\ Phys.\  B {\bf 195}, 481 (1982).
  %%CITATION = NUPHA,B195,481;%%
}

%\MaldacenaRE
\lref\AdSCFT{
  J.~M.~Maldacena,
  %``The large N limit of superconformal field theories and supergravity,''
  Adv.\ Theor.\ Math.\ Phys.\  {\bf 2}, 231 (1998)
  [Int.\ J.\ Theor.\ Phys.\  {\bf 38}, 1113 (1999)]
  [arXiv:hep-th/9711200].
  %%CITATION = IJTPB,38,1113;%%
 E.~Witten,
  %``Anti-de Sitter space and holography,''
  Adv.\ Theor.\ Math.\ Phys.\  {\bf 2}, 253 (1998)
  [arXiv:hep-th/9802150].
  %%CITATION = 00203,2,253;%%
 S.~S.~Gubser, I.~R.~Klebanov and A.~M.~Polyakov,
  %``Gauge theory correlators from non-critical string theory,''
  Phys.\ Lett.\  B {\bf 428}, 105 (1998)
  [arXiv:hep-th/9802109].
  %%CITATION = PHLTA,B428,105;%%
}

%\AlishahihaMD
\lref\AlishahihaMD{
  M.~Alishahiha, A.~Karch, E.~Silverstein and D.~Tong,
  %``The dS/dS correspondence,''
  AIP Conf.\ Proc.\  {\bf 743}, 393 (2005)
  [arXiv:hep-th/0407125].
  %%CITATION = APCPC,743,393;%%
}

%% AdS/CFT on de Sitter .

%\CallanPY
\lref\CallanPY{
  C.~G.~Callan, Jr. and F.~Wilczek,
  %``On geometric entropy,''
Phys.\ Lett.\ B {\bf 333}, 55 (1994).
[hep-th/9401072].
%%CITATION = hep-th/9401072%%
}
%\HawkingDA
\lref\HawkingDA{
  S.~Hawking, J.~M.~Maldacena and A.~Strominger,
  %``DeSitter entropy, quantum entanglement and AdS/CFT,''
  JHEP {\bf 0105}, 001 (2001)
  [arXiv:hep-th/0002145].
  %%CITATION = JHEPA,0105,001;%%
}

%\BuchelIU
\lref\BuchelIU{
  A.~Buchel and A.~A.~Tseytlin,
  %``Curved space resolution of singularity of fractional D3-branes on
  %conifold,''
  Phys.\ Rev.\  D {\bf 65}, 085019 (2002)
  [arXiv:hep-th/0111017].
  %%CITATION = PHRVA,D65,085019;%%
}

%\BuchelWF
\lref\BuchelWF{
  A.~Buchel,
  %``Gauge / gravity correspondence in accelerating universe,''
  Phys.\ Rev.\  D {\bf 65}, 125015 (2002)
  [arXiv:hep-th/0203041].
  %%CITATION = PHRVA,D65,125015;%%
}
%\BuchelTJ
\lref\BuchelTJ{
  A.~Buchel, P.~Langfelder and J.~Walcher,
  %``Does the tachyon matter?,''
  Annals Phys.\  {\bf 302}, 78 (2002)
  [arXiv:hep-th/0207235].
  %%CITATION = APNYA,302,78;%%
}
%\HirayamaJN
\lref\HirayamaJN{
  T.~Hirayama,
  %``A holographic dual of CFT with flavor on de Sitter space,''
  JHEP {\bf 0606}, 013 (2006)
  [arXiv:hep-th/0602258].
  %%CITATION = JHEPA,0606,013;%%
}

%\BuchelEM
\lref\BuchelEM{
  A.~Buchel,
  %``Inflation on the resolved warped deformed conifold,''
  Phys.\ Rev.\  D {\bf 74}, 046009 (2006)
  [arXiv:hep-th/0601013].
  %%CITATION = PHRVA,D74,046009;%%
}

%\MaldacenaVR
\lref\MaldacenaVR{
  J.~M.~Maldacena,
  %``Non-Gaussian features of primordial fluctuations in single field inflationary models,''
JHEP {\bf 0305}, 013 (2003).
[astro-ph/0210603].
%%CITATION = astro-ph/0210603%%
}
%\HarlowKE
\lref\HarlowKE{
  D.~Harlow and D.~Stanford,
  %``Operator Dictionaries and Wave Functions in AdS/CFT and dS/CFT,''
[arXiv:1104.2621 [hep-th]].
%%CITATION = arXiv:1104.2621%%
}

%\BuchelIU
\lref\BuchelIU{
  A.~Buchel and A.~A.~Tseytlin,
  %``Curved space resolution of singularity of fractional D3-branes on conifold,''
Phys.\ Rev.\ D {\bf 65}, 085019 (2002).
[hep-th/0111017].
%%CITATION = hep-th/0111017%%
}
%\BuchelWF
\lref\BuchelWF{
  A.~Buchel,
  %``Gauge / gravity correspondence in accelerating universe,''
Phys.\ Rev.\ D {\bf 65}, 125015 (2002).
[hep-th/0203041].
%%CITATION = hep-th/0203041%%
}

%\AmicoAG
\lref\AmicoAG{
  L.~Amico, R.~Fazio, A.~Osterloh and V.~Vedral,
  %``Entanglement in many-body systems,''
Rev.\ Mod.\ Phys.\  {\bf 80}, 517 (2008).
[quant-ph/0703044 [QUANT-PH]].
%%CITATION = quant-ph/0703044%%
}

%\HorodeckiZZ
\lref\HorodeckiZZ{
  R.~Horodecki, P.~Horodecki, M.~Horodecki and K.~Horodecki,
  %``Quantum entanglement,''
Rev.\ Mod.\ Phys.\  {\bf 81}, 865 (2009).
[quant-ph/0702225].
%%CITATION = quant-ph/0702225%%
}

%\CasiniSR
\lref\CasiniSR{
  H.~Casini and M.~Huerta,
  %``Entanglement entropy in free quantum field theory,''
J.\ Phys.\ A {\bf 42}, 504007 (2009).
[arXiv:0905.2562 [hep-th]].
%%CITATION = arXiv:0905.2562%%
}

%\NgXP
\lref\NgXP{
  G.~S.~Ng and A.~Strominger,
  %``State/Operator Correspondence in Higher-Spin dS/CFT,''
[arXiv:1204.1057 [hep-th]].
%%CITATION = arXiv:1204.1057%%
}

%\LiuEEA
\lref\LiuEEA{
  H.~Liu and M.~Mezei,
  %``A Refinement of entanglement entropy and the number of degrees of freedom,''
[arXiv:1202.2070 [hep-th]].
%%CITATION = arXiv:1202.2070%%
}

%\SolodukhinGN
\lref\SolodukhinGN{
  S.~N.~Solodukhin,
  %``Entanglement entropy of black holes,''
Living Rev.\ Rel.\  {\bf 14}, 8 (2011).
[arXiv:1104.3712 [hep-th]].
%%CITATION = arXiv:1104.3712%%
}

%\NishiokaUN
\lref\NishiokaUN{
  T.~Nishioka, S.~Ryu and T.~Takayanagi,
  %``Holographic Entanglement Entropy: An Overview,''
J.\ Phys.\ A {\bf 42}, 504008 (2009).
[arXiv:0905.0932 [hep-th]].
%%CITATION = arXiv:0905.0932%%
}

%\HartleAI
\lref\HartleAI{
  J.~B.~Hartle and S.~W.~Hawking,
  %``Wave Function of the Universe,''
Phys.\ Rev.\ D {\bf 28}, 2960 (1983)..
%%CITATION = PRINT-83-0937 (CAMBRIDGE)%%
}

%\BunchYQ
\lref\BunchYQ{
  T.~S.~Bunch and P.~C.~W.~Davies,
  %``Quantum Field Theory in de Sitter Space: Renormalization by Point Splitting,''
Proc.\ Roy.\ Soc.\ Lond.\ A {\bf 360}, 117 (1978)..
}

%\ChernikovZM
\lref\ChernikovZM{
  N.~A.~Chernikov and E.~A.~Tagirov,
  %``Quantum theory of scalar fields in de Sitter space-time,''
Annales Poincare Phys.\ Theor.\ A {\bf 9}, 109 (1968)..
}

%\SolodukhinDH
\lref\SolodukhinDH{
  S.~N.~Solodukhin,
  %``Entanglement entropy, conformal invariance and extrinsic geometry,''
Phys.\ Lett.\ B {\bf 665}, 305 (2008).
[arXiv:0802.3117 [hep-th]].
%%CITATION = arXiv:0802.3117%%
}

%\CallanPY
\lref\CallanPY{
  C.~G.~Callan, Jr. and F.~Wilczek,
  %``On geometric entropy,''
Phys.\ Lett.\ B {\bf 333}, 55 (1994).
[hep-th/9401072].
%%CITATION = hep-th/9401072%%
}

%\WittenZW
\lref\WittenZW{
  E.~Witten,
  %``Anti-de Sitter space, thermal phase transition, and confinement in gauge theories,''
Adv.\ Theor.\ Math.\ Phys.\  {\bf 2}, 505 (1998).
[hep-th/9803131].
%%CITATION = hep-th/9803131%%
}

%\RyuBV
\lref\RyuBV{
  S.~Ryu and T.~Takayanagi,
  %``Holographic derivation of entanglement entropy from AdS/CFT,''
Phys.\ Rev.\ Lett.\  {\bf 96}, 181602 (2006).
[hep-th/0603001].
%%CITATION = hep-th/0603001%%
}

%\HubenyXT
\lref\HubenyXT{
  V.~E.~Hubeny, M.~Rangamani and T.~Takayanagi,
  %``A Covariant holographic entanglement entropy proposal,''
JHEP {\bf 0707}, 062 (2007).
[arXiv:0705.0016 [hep-th]].
%%CITATION = arXiv:0705.0016%%
}

%\MaldacenaUN
\lref\MaldacenaUN{
  J.~Maldacena,
  %``Vacuum decay into Anti de Sitter space,''
[arXiv:1012.0274 [hep-th]].
%%CITATION = arXiv:1012.0274%%
}
%\BombelliRWFZW
\lref\BombelliRW{
  L.~Bombelli, R.~K.~Koul, J.~Lee and R.~D.~Sorkin,
  %``A Quantum Source of Entropy for Black Holes,''
Phys.\ Rev.\ D {\bf 34}, 373 (1986)..
%%CITATION = PRINT-86-0371 (SYRACUSE)%%
}

%\SrednickiIM
\lref\SrednickiIM{
  M.~Srednicki,
  %``Entropy and area,''
Phys.\ Rev.\ Lett.\  {\bf 71}, 666 (1993).
[hep-th/9303048].
%%CITATION = hep-th/9303048%%
}

%\HertzbergUV
\lref\HertzbergUV{
  M.~P.~Hertzberg and F.~Wilczek,
  %``Some Calculable Contributions to Entanglement Entropy,''
Phys.\ Rev.\ Lett.\  {\bf 106}, 050404 (2011).
[arXiv:1007.0993 [hep-th]].
%%CITATION = arXiv:1007.0993%%
}

%\SasakiYT
\lref\SasakiYT{
  M.~Sasaki, T.~Tanaka and K.~Yamamoto,
  %``Euclidean vacuum mode functions for a scalar field on open de Sitter space,''
Phys.\ Rev.\ D {\bf 51}, 2979 (1995).
[gr-qc/9412025].
%%CITATION = gr-qc/9412025%%
}
%\BucherGB
\lref\BucherGB{
  M.~Bucher, A.~S.~Goldhaber and N.~Turok,
  %``An open universe from inflation,''
Phys.\ Rev.\ D {\bf 52}, 3314 (1995).
[hep-ph/9411206].
%%CITATION = hep-ph/9411206%%
}

%\BytsenkoBC
\lref\BytsenkoBC{
  A.~A.~Bytsenko, G.~Cognola, L.~Vanzo and S.~Zerbini,
  %``Quantum fields and extended objects in space-times with constant curvature spatial section,''
Phys.\ Rept.\  {\bf 266}, 1 (1996).
[hep-th/9505061].
%%CITATION = UTF-325%%
}
%\CasiniKT
\lref\Entsph{
  J.~S.~Dowker,
  %``Hyperspherical entanglement entropy,''
J.\ Phys.\ A {\bf 43}, 445402 (2010).
[arXiv:1007.3865 [hep-th]].
%%CITATION = arXiv:1007.3865%%
  J.~S.~Dowker,
  %``Entanglement entropy for even spheres,''
[arXiv:1009.3854 [hep-th]] and
%%CITATION = arXiv:1009.3854%%
  %``Entanglement entropy for odd spheres,''
[arXiv:1012.1548 [hep-th]].
%%CITATION = arXiv:1012.1548%%
}

%\CasiniKT
\lref\CasiniKT{
  H.~Casini and M.~Huerta,
  %``Entanglement entropy for the n-sphere,''
Phys.\ Lett.\ B {\bf 694}, 167 (2010).
[arXiv:1007.1813 [hep-th]].
%%CITATION = arXiv:1007.1813%%
}

%\KlebanovUF
\lref\KlebanovUF{
  I.~R.~Klebanov, S.~S.~Pufu, S.~Sachdev and B.~R.~Safdi,
  %``Renyi Entropies for Free Field Theories,''
JHEP {\bf 1204}, 074 (2012).
[arXiv:1111.6290 [hep-th]].
%%CITATION = arXiv:1111.6290%%
}

%\BanadosDF
\lref\BanadosDF{
  M.~Banados,
  %``Constant curvature black holes,''
Phys.\ Rev.\ D {\bf 57}, 1068 (1998).
[gr-qc/9703040].
%%CITATION = gr-qc/9703040%%
}

\lref\BanadosGM{
 M.~Banados, A.~Gomberoff and C.~Martinez,
  %``Anti-de Sitter space and black holes,''
Class.\ Quant.\ Grav.\  {\bf 15}, 3575 (1998).
[hep-th/9805087].
%%CITATION = hep-th/9805087%%
}
%\CasiniKV
\lref\CasiniKV{
  H.~Casini, M.~Huerta and R.~C.~Myers,
  %``Towards a derivation of holographic entanglement entropy,''
JHEP {\bf 1105}, 036 (2011).
[arXiv:1102.0440 [hep-th]].
%%CITATION = arXiv:1102.0440%%
}

%\ColemanAW
\lref\ColemanAW{
  S.~R.~Coleman and F.~De Luccia,
  %``Gravitational Effects on and of Vacuum Decay,''
Phys.\ Rev.\ D {\bf 21}, 3305 (1980)..
%%CITATION = SLAC-PUB-2463%%
}

%\HawkingDH
\lref\HawkingDH{
  S.~W.~Hawking and D.~N.~Page,
  %``Thermodynamics of Black Holes in anti-De Sitter Space,''
Commun.\ Math.\ Phys.\  {\bf 87}, 577 (1983)..
%%CITATION = PRINT-83-0019 (CAMBRIDGE)%%
}

%\HertogRZ
\lref\HertogRZ{
  T.~Hertog and G.~T.~Horowitz,
  %``Towards a big crunch dual,''
JHEP {\bf 0407}, 073 (2004).
[hep-th/0406134].
%%CITATION = hep-th/0406134%%
%\HertogHU
  T.~Hertog and G.~T.~Horowitz,
  %``Holographic description of AdS cosmologies,''
JHEP {\bf 0504}, 005 (2005).
[hep-th/0503071].
%%CITATION = hep-th/0503071%%
}

%\AbbottKR
\lref\AbbottKR{
  L.~F.~Abbott and S.~R.~Coleman,
  %``The Collapse Of An Anti-de Sitter Bubble,''
Nucl.\ Phys.\ B {\bf 259}, 170 (1985)..
%%CITATION = HUTP-85-A021%%
}

%\LiuEEA
\lref\LiuEEA{
  H.~Liu and M.~Mezei,
  %``A Refinement of entanglement entropy and the number of degrees of freedom,''
[arXiv:1202.2070 [hep-th]].
%%CITATION = arXiv:1202.2070%%
}

%\AbajoArrastiaYT
\lref\AbajoArrastiaYT{
  J.~Abajo-Arrastia, J.~Aparicio and E.~Lopez,
  %``Holographic Evolution of Entanglement Entropy,''
JHEP {\bf 1011}, 149 (2010).
[arXiv:1006.4090 [hep-th]].
%%CITATION = arXiv:1006.4090%%
}

%%%%%%%%%%%%%%%%%%%%%%%%%%%%%%%%%%%%%%%%%%%%%%%%%%%%%%%%%%%%%%%%%%%%
\rightline{PUPT-2428}
\Title{\vbox{\baselineskip12pt \hbox{} \hbox{
} }} {\vbox{\centerline{ Entanglement entropy in de Sitter space
  }
\centerline{
  }
}}
\bigskip
\centerline{  Juan Maldacena$^*$ and Guilherme L. Pimentel$^\dagger$}
\bigskip
\centerline{ \it $^*$ School of Natural Sciences, Institute for
Advanced Study} \centerline{\it Princeton, NJ 08540, USA}
\centerline{ \it $^\dagger$ Joseph Henry Laboratories, Princeton University}
\centerline{\it Princeton, NJ 08544, USA}

\vskip .3in \noindent
%%%%%%%%%%%%%%%%%%%%%%%%%%%%%%%%%%%%%%%%%%%%%%%%%%%%%%%%%%%%%%%%%%%%%%%%%%%%%%%%%%%%%%%%%%%%

We compute the entanglement entropy for some quantum field theories on de Sitter space.
We consider a superhorizon size
spherical surface that divides the spatial slice into two regions, with
the field theory in the standard vacuum state. First, we study a free massive scalar field.
Then, we consider a strongly coupled field theory with a gravity dual,  computing the entanglement
using the  gravity solution. In even dimensions,
the interesting piece of the entanglement entropy is proportional
to the number of e-foldings that elapsed since the spherical region was inside the horizon. In odd
dimensions it is contained in a certain finite piece. In both cases the entanglement captures the
long range correlations produced by the expansion.

%%%%%%%%%%%%%%%%%%%%%%%%%%%%%%%%%%%%%%%%%%%%%%%%%%%%%%%%%%%%%%%%%%%%%%%%%%%%%%%%%%%%%%%%%%%%

 \Date{ }

%%%%%%%%%%%%%%%%%%%%%%%%%%%%%%%%%%%%%%%%%%%%%%%%%%%%%%%%%%%%%%%%%%%%%%%%%

\newsec{Introduction}

Entanglement entropy is a useful tool to characterize states with long range quantum order in
condensed matter physics (see \refs{\AmicoAG,\HorodeckiZZ} and references therein). It is also useful in quantum field theory to characterize the nature
of the long range correlations that we have in the vacuum (see e.g. \refs{\BombelliRW,\SrednickiIM,\CasiniSR} and
references therein).

We   study the entanglement entropy for quantum field theories in de Sitter space. We  choose the standard
vacuum  state  \refs{\BunchYQ,\ChernikovZM,\HartleAI} (the Euclidean,  Hartle-Hawking/Bunch-Davies/Chernikov-Tagirov vacuum).
We do not include dynamical gravity. In particular, the entropy we compute should not be confused
with the gravitational de Sitter entropy.

Our motivation is to quantify the degree of superhorizon correlations that are generated by the
cosmological expansion.

We consider a spherical surface that divides the spatial slice into the interior and exterior. We
compute the entanglement entropy  by
tracing over the exterior.   We   take the size of this sphere, $R$, to be much bigger than the de Sitter radius, $R\gg R_{dS} = H^{-1}$,
 where $H$ is Hubble's constant. Of course, for  $R\ll R_{dS}$ we expect  the same result as in flat space.
If $R = R_{dS}$, then we would have the usual thermal density matrix in the static patch and its associated
entropy\foot{This can be regarded as a (UV divergent)  ${\cal O}(G_N^0)$ correction  to the gravitational entropy 
of de Sitter space \CallanPY .}. 
As usual, the entanglement entropy has a UV divergent contribution which we   ignore,  since it comes
from local physics. For very large spheres, and in four dimensions,
the finite piece has a term that goes like the area of the sphere and one that goes like the logarithm
of the area. We focus on the coefficient of the logarithmic piece. In odd spacetime dimensions there
are finite terms that go like positive powers of the area and a constant term. We then focus on
the constant term.

We first calculate the entanglement
entropy  for a  free massive scalar field. To determine it, one needs to find the density matrix from tracing out the degrees of freedom outside of the surface.
 When the spherical surface is taken all the way to the boundary of de Sitter space the problem develops
 an  $SO(1,3)$ symmetry. This symmetry is very helpful for
computing  the density matrix and the associated entropy. Since we have the density matrix, it is also easy
to compute the R\'enyi entropies.

We then study the entanglement entropy of field theories with a gravity dual. When the dual is known, we use the proposal of  \refs{\RyuBV,\HubenyXT} to calculate the entropy. It boils down to an extremal area problem.
 The answer for the entanglement entropy depends drastically on the properties of the gravity dual. In particular,
if the gravity dual has a hyperbolic Friedman-Robertson-Walker spacetime inside, then there is a non-zero
contribution at order $N^2$ for the ``interesting'' piece of the entanglement entropy. Otherwise, the order
$N^2$ contribution vanishes.

This  provides some further hints that the FRW region is indeed somehow contained
in the field theory in de Sitter space \MaldacenaUN . More precisely, it should be
contained in the superhorizon correlations of colored fields\foot{A holographic calculation of the entanglement entropy associated to a quantum quench is presented in \AbajoArrastiaYT . A quantum quench is the sudden perturbation of a pure state. The subsequent relaxation back to equilibrium can be understood in terms of the entanglement entropy of the quenched region. There, one has a contribution to the (time dependent)  entropy coming from the region behind the horizon of the holographic dual. }.

The paper is organized as follows. In section~2, we discuss general features of entanglement entropy in de Sitter. In section~3, we  consider  a free scalar field and compute its entanglement entropy. In section~4, we write holographic duals of field theories in de Sitter, and compute the entropy of spherical surfaces in these theories. We end with a discussion.
Some more technical details are presented in the appendices.

\newsec{General features of entanglement entropy in de Sitter}

Entanglement entropy is defined as follows \CallanPY. At some given time slice, we consider a closed surface $\Sigma$
which separates the  slice into a region inside the surface and a region outside. In a local quantum field
theory we expect to have an approximate decomposition of the Hilbert space into $H = H_{in} \times H_{out}$
where $H_{in}$ contains modes localized inside the surface and $H_{out}$ modes localized outside.
One can then define a density matrix $\rho_{in} = \Tr_{H_{out}} |\psi \rangle \langle \psi|$ obtained by tracing over the outside Hilbert space. The entanglement entropy is the von Neumann entropy obtained from this density matrix:
\eqn\entr{
S=-\Tr \rho_{in} \log \rho_{in}
}

\subsec{Four dimensions}

We consider de Sitter space in the flat slicing
\eqn\poinc{
ds^2={1 \over (H \eta)^2} ( -d \eta^2 + dx_1^2+dx_2^2+dx_3^2)
}
where $H$ is the Hubble scale and $\eta$ is conformal time.
We   consider surfaces that sit at constant $\eta$ slices.  We consider a free, minimally coupled,
scalar field of mass $m$ in the usual vacuum state \refs{\BunchYQ,\ChernikovZM,\HartleAI}.

As in any quantum field theory, the entanglement entropy is UV divergent
\eqn\entropy{\eqalign{
&S=S_{\rm UV-divergent}+S_{\rm UV-finite}
}} The UV divergencies are
due to local effects and have the the form
\eqn\entuv{
S_{\rm UV-divergent}=c_1{A\over \epsilon^2} + \log(\epsilon H)(c_2 + c_3 Am^2 +c_4 A H^2)
}
where $\epsilon$ is the $UV$ cutoff. The first term is the well known area contribution to the entropy \refs{\BombelliRW,\SrednickiIM}, coming from entanglement of particles close to the surface considered.
The logarithmic terms involving $c_2$ and $c_3$  also arise in flat space. Finally, the last term
 involves the curvature of the bulk space\foot{
In de Sitter there is only one curvature scale, but in general we could write terms as
\eqn\genflog{
S_{\log \epsilon H} = \int_{\Sigma}\left(a R_{\mu\nu\rho\sigma} n^\mu_i n^\rho_i n^\nu_j n^\sigma_j + b R_{\mu\nu} n^\mu_i n^\nu_i + c R + d K^{\mu\nu}_i K_{i\mu\nu} + e K^\mu_{i\mu} K^\nu_{i\nu} + \cdots \right)
}
where $K$ are the extrinsic curvatures and $i,j$ label the two  normal directions
and $\mu, \nu, \cdots$ are spacetime indices The extrinsic curvatures also contribute to $c_2$ in \entuv .  One could also write a term that depends on the intrinsic curvature of the surface, $R_\Sigma$, but the Gauss-Codazzi relations can be used to relate it to the other terms in \genflog.
}. All these UV divergent terms arise
from local effects and their coefficients are the same as what we would have obtained in flat space.
We have included $H$ as a scale inside the logarithm.
This is just an arbitrary definition, we could also have used $m$ \HertzbergUV, when $m$ is non-zero.

Our focus  is on the UV finite terms that contain information about the long range correlations of
the quantum state in de Sitter space.
The entropy is invariant under the isometries of $dS$. This is true for both pieces in \entropy .
In addition, we expect that  the long distance part of the state becomes time independent.
 More precisely, the long range entanglement  was
established when these distances were subhorizon size. Once they moved outside the horizon
we do not expect to be able to modify this entanglement by subsequent evolution.
Thus, we expect that the long range part of the entanglement entropy should be constant as we go
to late times. So, if we fix a surface in comoving $x$ coordinates in \poinc , and
we keep this surface fixed as we move to late times, $\eta \to 0$, then we naively expect that
the entanglement should be constant. This expectation is not quite right because new modes are
coming in at late times. However, all these modes only give rise to entanglement at short distances in
comoving coordinates. The effects of this entanglement could be written in a local fashion.

In conclusion, we expect that the UV-finite piece of the entropy is given by

\eqn\entir{
S_{\rm UV-Finite }= c_5 A H^2 +{  c_6 \over 2 }  \log( A H^2)  + {\rm finite}  = c_5 {A_c \over \eta^2} + c_6 \log\eta
+ {\rm finite} }
where $A$ is the proper area of the surface and $A_c$ is the area in comoving coordinates $(A = {A_c \over
H^2 \eta^2 })$. The finite piece is a bit ambiguous due to the presence of the logarithmic term.

The coefficient of the logarithmic term, $c_6$, contains information about the long range entanglement
of the state. This term looks similar to the UV divergent logarithmic term in \entuv , but they should
not be confused with each other. If we had a conformal field theory in de Sitter they would be equal.
However, in a non-conformal theory they are not equal ($c_6 \not = c_2$).
 For general surfaces, the coefficient of the logarithm will
depend on two combinations of  the extrinsic curvature of the surface in comoving coordinates. For simplicity we consider  a
sphere here\foot{
It is enough to do the computation for another surface, say a cylinder, to determine the
second coefficient and  have a result that is
valid for general surfaces \SolodukhinDH . In other words, for a general surface we
have $c_6 = f_1 \int K_{ab} K_{ab} + f_2 \int (K_{aa})^2 $ where $f_1,~f_2$ are some constants and
$K_{ab}$ is the extrinsic curvature of the surface within the spatial slice.}. This general form of the entropy, \entir , will be
confirmed by our explicit computations below. 

We define the ``interesting'' part of the 
entropy to be the coefficient of the logarithm, $S_{\rm intr} \equiv c_6$. The $UV$-finite area term, with coefficient $c_5$, though physically interesting, is not easily  calculable with our method. It receives 
contributions from the entanglement at distances of a few Hubble radii from the entangling surface. 
 It would be nice to find a way to isolate this contribution and compute $c_5$ exactly. We could only do that in the case where the theory has a gravity dual.

\subsec{Three dimensions}

For three dimensional de Sitter space we can have  a similar discussion.
\eqn\entthree{\eqalign{
&S_{}=d_1 {A\over \epsilon}  +S_{\rm UV-finite} \cr
&S_{\rm UV-finite}= d_2 {A H } + d_3  = d_2  {A_c \over \eta} + d_3
}}
Here there is no logarithmic term. The interesting term is $d_3$ which is the finite piece.
So we define $S_{\rm intr} \equiv d_3$.

A similar discussion exists in all other dimensions. For even spacetime dimensions the interesting term
is the logarithmic one and for odd dimensions it is the constant.
One can isolate these interesting terms by taking appropriate derivatives with respect to the physical
area, as done in \LiuEEA\ in a similar context\foot{See formula (1.1) of \LiuEEA .}.

Note that we are considering quantum fields in a fixed spacetime. We have no gravity. And we are making
no contact with the gravitational de Sitter entropy which is the area of the horizon in Planck units.

\newsec{Entanglement entropy for a free massive scalar field in de Sitter}

Here we compute the entropy of a free massive scalar field for a spherical entangling surface.

\subsec{Setup of the problem}

Consider, in flat coordinates, a spherical surface $S^2$ defined by $x_1^2+x_2^2+x_3^2=R_c^2$. We consider $R_c\gg \eta$. This means that the surface is much bigger than the horizon.

\ifig\setup{
Setup of the problem: (a) We consider a sphere with radius much greater than the horizon size, at late conformal time $\eta$, in flat slices. (b) This problem can be mapped to half of a 3-sphere $S^3$, also with boundary $S^2$, but now the equator, at late global time $\tau_B$. (c) We can also  describe this problem
using  hyperbolic slices. The interior of the sphere maps to
the
  ``left" (L) hyperbolic slice. The Penrose diagrams for all situations are depicted below the geometric sketches.} {\epsfxsize5in\epsfbox{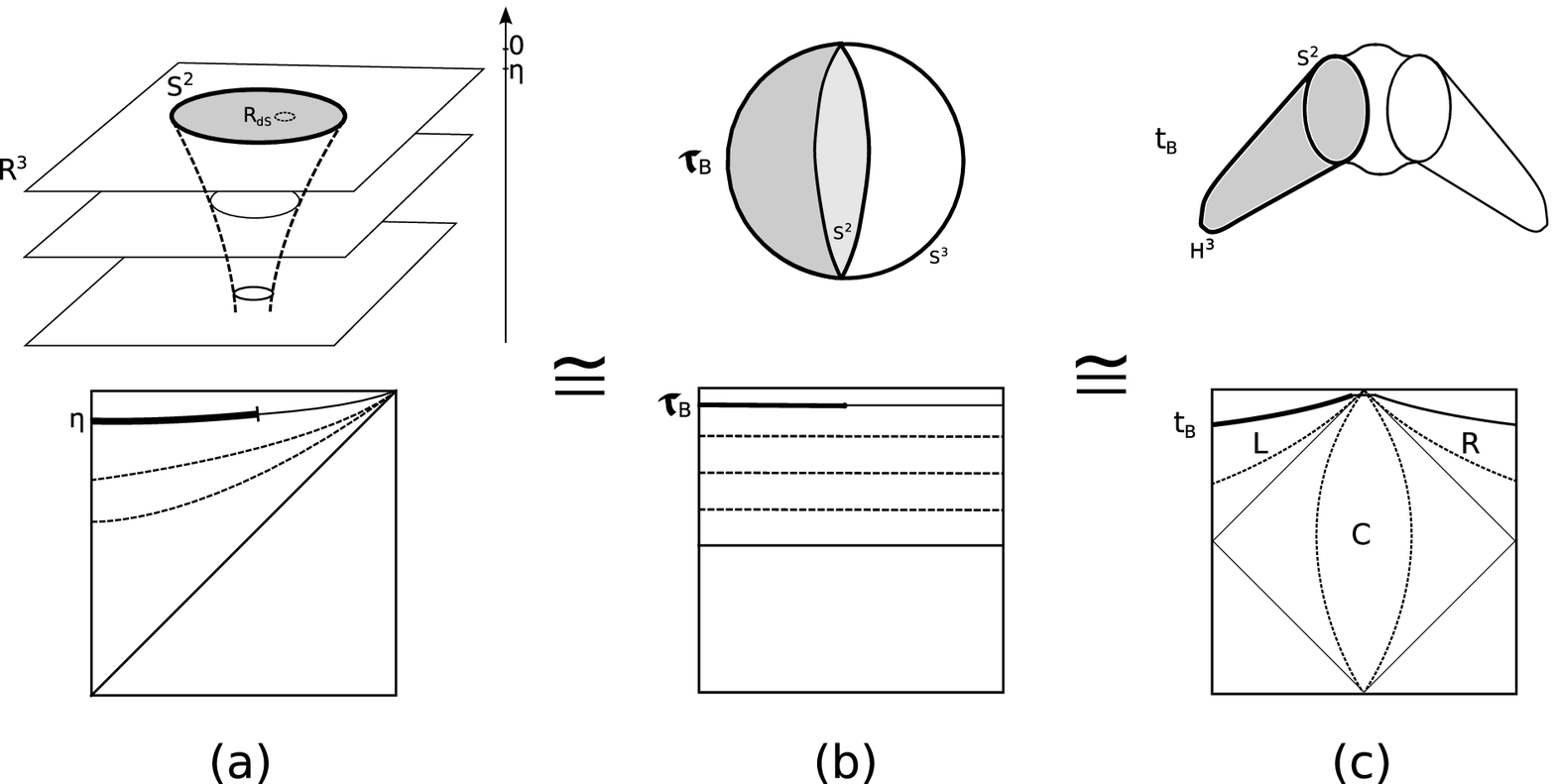}}

If we could neglect the $\eta$ dependent terms, we can take the limit $\eta \to 0$, keeping $R_c$ fixed.
This then becomes a surface on the boundary. This surface is left invariant by an $SO(1,3)$ subgroup
of the $SO(1,4)$ de Sitter isometry group. We expect that the coefficient of the logarithmic term that
we discussed above is also invariant under this group. It is therefore convenient to choose a
coordinate system where $SO(1,3)$ is realized more manifestly. This is done in two steps. First we
can consider de Sitter in global coordinates, where  the equal time slices are three-spheres.
Then we can choose the entangling surface to be the two-sphere equator of the three-sphere.
 In fact, at $\eta=0$, we can certainly map any two sphere on the boundary of de Sitter to the equator
of $S^3$ by a de Sitter isometry. Finally, to regularize this problem we can then move back the two
sphere to a very late fixed global time surface.

We can then choose a coordinate system where the SO(1,3) symmetry is  realized geometrically in
a simple way. Namely, this SO(1,3) is the symmetry group acting on
 hyperbolic slices in some coordinate system
that we describe below.

\subsec{Wavefunctions of free fields in hyperbolic slices and the Euclidean vacuum}

The hyperbolic/open slicing of  de Sitter
 space was studied in detail in \refs{\BucherGB, \SasakiYT}. It can be obtained by analytic continuation of the sphere $S^4$ metric, sliced by $S^3$s. The $S^4$ is described in embedding coordinates by $X_1^2+...+X_5^2=H^{-2}$. The coordinates are parametrized by angles in the following way:
\eqn\angsfo{
X_5 = H^{-1}\cos \tau_E \cos \rho_E,~~~~~~X_4=H^{-1}\sin \tau_E, ~~~~~~
X_{1,2,3}=H^{-1}\cos \tau_E \sin\rho_E n_{1,2,3}
}
where $n_i$ are the components of a unit vector in $R^3$.  The metric in Euclidean signature is given by:
\eqn\meteuc{
ds_E^2=H^{-2}(d\tau_E^2+\cos^2\tau_E(d\rho_E^2+\sin^2\rho_E ~d\Omega_2^2))
}
We analytically continue $X_5 \to i X_0$. Then the Lorentzian manifold is divided in three parts, related to the Euclidean coordinates by:
\eqn\anaco{\eqalign{
&R: ~~\cases{\tau_E= { \pi \over 2 } - i t_R   & $t_R \ge 0 $\cr
	\rho_E = - i r_R & $r_R \ge 0$	}\cr
&C:~~\cases{\tau_E = \tau_C & $ -\pi/2 \le t_C \le \pi/2$ \cr
		\rho_E = {\pi \over 2} - i r_C & $ -\infty < r_C < \infty$}\cr
&L:~~\cases{ \tau_E = - {\pi \over 2 } + i t_L  & $t_L \ge 0 $\cr
	\rho_E= - i r_L & $r_L \ge 0$}
}}

The metric in each region is given by:
\eqn\methyp{\eqalign{
&ds_R^2=H^{-2}(-dt_R^2+\sinh^2 t_R(dr_R^2+\sinh^2 r_R d\Omega_2^2))\cr
&ds_C^2=H^{-2}(dt_C^2+\cos^2 t_C(-dr_C^2+\cosh^2 r_C d\Omega_2^2))\cr
&ds_L^2=H^{-2}(-dt_L^2+\sinh^2 t_L(dr_L^2+\sinh^2 r_L d\Omega_2^2))
}}
We now consider a minimally coupled\foot{If we had a  coupling to the scalar curvature $\xi R \phi^2$, we
can simply shift the mass   $m_{eff}^2=m^2+6 \xi H^2$ and consider the minimally coupled one.}
massive scalar field in $dS_4$, with action given by $S=
{1 \over 2}\int \sqrt{-g} ( -(\nabla \phi)^2 - m^2 \phi^2)$. The equations of motion for the mode functions
in  the $R$ or $L$ regions are
\eqn\eomhyp{
\left[ {1 \over \sinh^3 t} {\partial \over \partial t} \sinh^3 t {\partial \over \partial t} - {1 \over \sinh^2 t} {\bf L_{H^3}^2} + {9 \over 4} - \nu^2 \right] u(t,r,\Omega) = 0
}
Where ${\bf L_{H^3}^2}$ is the Laplacian in the unit hyperboloid, and the parameter $\nu$ is
\eqn\mass{
\nu = \sqrt{ {9 \over 4} - {m^2 \over H^2}}
}
When $\nu ={ 1 \over 2}$ (or ${ m^2 \over H^2}  = 2$) we have a conformally coupled massless scalar. In this
case we should recover the flat space answer for the entanglement entropy, since de Sitter is
conformally flat.
We will consider first situations where ${ m^2 \over H^2} \geq 2$, so that $0\le\nu\le1/2$ or $\nu$ imaginary.
 The minimally coupled massless case corresponds to $\nu= 3/2$. We will later comment on the low mass region,
${ m^2 \over H^2} < 2$ or  $1/2 < \nu \le 3/2$.

The wavefunctions are labeled by quantum numbers corresponding to the Casimir on $H^3$ and angular momentum on $S^2$:
\eqn\waved{
u_{plm}\sim {H \over \sinh t} \chi_p(t) Y_{plm}(r, \Omega_2)~,~~~~~~~ - {\bf L_{H^3}} Y_{plm} = (1 + p^2) Y_{plm}
}
The $Y_{plm}$ are eigenfunctions on the hyperboloid, analogous to the standard spherical harmonics. Their expressions can be found in \SasakiYT . 

The time dependence (other than the $1/\sinh t$ factor) is contained in the functions $\chi_p(t)$. The equation of motion \eomhyp\ is a Legendre equation and the solutions are given in terms of
Legendre functions $P_a^b(x)$.
In order to pick the ``positive frequency'' wavefunctions corresponding to the Euclidean vacuum
we need to demand that they are
analytic when they are continued to the lower hemisphere. These wavefunctions have support
on both the Left and Right regions. This gives \SasakiYT
\eqn\wavef{
\chi_{p,\sigma}=\cases{ \displaystyle{1 \over 2 \sinh \pi p} \left( {e^{\pi p} - i \sigma e^{-i \pi \nu} \over \Gamma(\nu+ i p +1/2)} P^{ip}_{\nu-1/2}(\cosh t_R) -  {e^{-\pi p} - i \sigma e^{-i \pi \nu} \over \Gamma(\nu- i p +1/2)} P^{-ip}_{\nu-1/2}(\cosh t_R) \right) & \cr
  \displaystyle{\sigma \over 2 \sinh \pi p} \left( {e^{\pi p} - i \sigma e^{-i \pi \nu} \over \Gamma(\nu+ i p +1/2)} P^{ip}_{\nu-1/2}(\cosh t_L) -  {e^{-\pi p} - i \sigma e^{-i \pi \nu} \over \Gamma(\nu- i p +1/2)} P^{-ip}_{\nu-1/2}(\cosh t_L) \right)
}
}
The index $\sigma$ can take the values $\pm 1$.  For each $\sigma$ the top line gives the function on the
R hyperboloid and the bottom line gives the value of the function on the L hyperboloid.  There are two solutions (two values of $\sigma$) because we started from  two hyperboloids.

The field operator is written in terms of these mode functions as
\eqn\field{
\hat\phi(x) =\int dp \sum_{\sigma,l,m} (a_{  \sigma p l m} u_{ \sigma p lm}(x) + a^\dagger_{ \sigma p l m} \overline u_{  \sigma p  l m}(x)) }

To trace out the degrees of freedom in, say, the $R$ space, we change basis to functions that have support on either the $R$ or $L$ regions. It does not matter which functions we choose to describe the Hilbert space.
The crucial simplification of this coordinate system is that
the entangling surface, when taken to the de Sitter boundary, preserves all the isometries of the
$H^3$ slices. This implies that the entanglement is diagonal in the $p,l,m$ indices since these are
all eigenvalues of some symmetry generator. Thus, to compute this entanglement we only need to
look at the analytic properties of \wavef\ for each value of $p$.

Let us first consider the case that $\nu$ is real.
For the R region we take  basis functions equal to  the Legendre functions $P^{ip}_{\nu-1/2}(\cosh t_R)$ and $P^{-ip}_{\nu-1/2}(\cosh t_R)$, and zero in the $L$ region. These are the positive and negative frequency
 wavefunctions in the R region.  We do the same in the $L$ region.
 These should be properly normalized with respect to the Klein-Gordon norm, which would yield a normalization factor $N_p$. We can write the original mode functions, \wavef ,  in terms of these new ones in matricial form:
\eqn\basis{\eqalign{
&\cases{\chi^\sigma=N^{-1}_p\sum_{q=R,L} (\alpha^\sigma_qP^q+\beta^\sigma_q\overline P^q) \cr\overline \chi^\sigma =N^{-1}_p\sum_{q=R,L} (\overline\beta^\sigma_qP^q+\overline\alpha^\sigma_q\bar P^q)} \Rightarrow \chi^I= M^I_JP^J  N^{-1}_p \cr
&\sigma=\pm1,~~P^{R,L}\equiv P^{ip}_{\nu-1/2}(\cosh t_{R,L}),~~ \chi^I\equiv\left(\matrix{\chi^\sigma \cr \overline\chi^\sigma}\right)
}}
The capital indices ($I,J$)
 run from 1 to 4, as we are grouping both the $\chi_\sigma$ and $\overline \chi_\sigma$.
The coefficients $\alpha$ and $\beta$ are simply the terms multiplying the corresponding $P$ functions in
\wavef , see appendix A for their explicit values.
 As the field operator should be the same under this change of basis, then it follows that:
\eqn\exp{\eqalign{
&\phi=a_I\chi^I=b_JP^J N^{-1}_p  \Rightarrow a_J=b_I (M^{-1})^I_J\cr
&M=\left(\matrix{\alpha & \beta \cr \overline\beta & \overline\alpha}\right),~~ M^{-1}=\left(\matrix{\gamma & \delta \cr \overline\delta & \overline\gamma}\right) \Rightarrow~~a_\sigma=\sum_{q=R,L}\gamma_{q\sigma} b_q+\overline\delta_{q\sigma} b^\dagger_q
}}
Here $a^I=(a_\sigma, a^\dagger_\sigma)$, $b^J = ( b_{L,R}, b^\dagger_{L,R})$, and $P^J = ( P_{L,R}, \bar P_{L,R} ) $. $M$ is a $2\times2$ matrix whose elements are $2\times 2$ matrices. The expression for
$M^{-1}$ is the definition of $\delta , ~\gamma $, etc.
The vacuum is defined so that $a_\sigma | \Psi \rangle = 0$. We want to write $|\Psi\rangle$ in terms of the $b_{R,L}$ oscillators and the vacua associated to each of these oscillators, $b_R|R\rangle =0$ and $b_L|L\rangle = 0$. As we are dealing with free fields, their Gaussian structure suggests the ansatz
\eqn\ansentb{
|\Psi \rangle = e^{{1 \over 2} \sum_{i,j=R,L}m_{ij} b^\dagger_i b^\dagger_j}|R\rangle |L\rangle
}
 and one can solve for $m_{ij}$ demanding that $a_\sigma |\Psi \rangle =0$. This gives
\eqn\msol{
 m_{ij} \gamma_{j \sigma} +\overline\delta_{i\sigma}=0 \Rightarrow m_{ij}=-\overline\delta_{i\sigma}(\gamma^{-1})_{\sigma j}
}
Using the expressions in \wavef\ (see appendix A)  we find for $m$:
\eqn\solex{
m_{ij} = e^{i \theta} {\sqrt{2} e^{-p \pi} \over \sqrt{\cosh 2 \pi p + \cos 2 \pi \nu}}\pmatrix{\cos \pi \nu & i \sinh p \pi \cr i \sinh p \pi & \cos \pi \nu}
}
Where $\theta$ is an unimportant phase factor, which can be absorbed in the definition of the $b^\dagger$ oscillators. In $m_{ij}$ the normalization factors $N_p$ drop out, so they never need to be computed.

The expression \ansentb , with \solex , needs to be simplified more before we can easily trace out the $R$
degrees of freedom. We would like to introduce new oscillators $c_L$ and $c_R$ (and their adjoints)
so that the original state $\Psi$ has the form
\eqn\finalc{
|\Psi \rangle = e^{\gamma c^\dagger_R c^\dagger_L} |R\rangle' |L \rangle'  }
 where $|R\rangle' |L \rangle'$
 are annihilated by $c_R, ~c_L$.
The details on the transformation are in appendix~A. Here we state the   result.
The $b$'s and $c$'s are related by:
\eqn\bogol{\eqalign{
& c_R=u b_R + v b^\dagger_R \cr
& c_L = \bar u b_L + \bar v b^\dagger_L, ~~~ |u|^2-|v|^2=1
}
}
Requiring that $c_R|\Psi\rangle = \gamma c^\dagger_L |\Psi \rangle$ and $c_L|\Psi \rangle = \gamma c_R^\dagger|\Psi \rangle$ imposes constraints on $u$ and $v$.
The system of equations has a solution with  $\gamma$  given by
\eqn\eigen{
\gamma = i {\sqrt{2} \over \sqrt{\cosh 2 \pi p + \cos 2 \pi \nu} + \sqrt{\cosh 2 \pi p + \cos 2 \pi \nu + 2}}
}

We have considered the case of $0 \le \nu \le 1/2$. For $\nu$ imaginary, \eigen\ is analytic under the substitution $\nu \to i \nu$, which corresponds to substituting $\cos 2 \pi \nu \to \cosh 2 \pi  i \nu$, so \eigen\ is also valid for this range of masses.  One can check directly, by redoing all the steps in the
above derivation, that the same final answer is obtained if we had assumed that $\nu$ was
purely imaginary.

\subsec{The density matrix }

 The full vacuum state is the product of the vacuum state for each oscillator. Each oscillator is labelled
by $p,l,m$. For each oscillator we can write the vacuum state as in \finalc .
Expanding \finalc\ and tracing over the right Hilbert space we get
\eqn\finre{
\rho_{p,l,m} = Tr_{H_R} ( |\Psi \rangle \langle \Psi | ) \propto
 \sum_{n=0}^\infty |\gamma_p|^2 |n ; p,l,m\rangle
\langle n;  p,l,m |
}
So, for given quantum numbers, the density matrix is diagonal. It takes  the form
$\rho_L(p) = (1-|\gamma_p|^2)  {\rm diag}(1, |\gamma_p|^2, |\gamma_p|^4, \cdots) $, normalized to $\Tr\rho_L=1$.
The full density matrix is simply the product of the density matrix for each value of $p, l , m$. This
reflects the fact that there is no entanglement among states with different $SO(1,3)$ quantum numbers. The density matrix for the conformally coupled case was computed before in 
\NgXP . 

Here, one can write the resulting density matrix as $\rho_L=e^{-\beta {\cal H}_{ent}}$ with ${\cal H}_{ent}$ called the entanglement hamiltonian. Here it seems natural to choose
 $\beta = 2 \pi$ as the inverse temperature of $dS$. Because the density matrix is diagonal, the entanglement Hamiltonian should be that of a gas of free particles, with the energy of each excitation a function of the $H^3$ Casimir and the mass of the scalar field. This does not appear to be related to any ordinary
dynamical Hamiltonian in de Sitter.
In other words, take $\rho_L  \propto 
{\rm diag} (1,|\gamma_p|^2,|\gamma_p|^4,...)$ then the entanglement Hamiltonian for each particles is
 $H_p =  E_p c^\dagger_p c_p$, with $E_p = -{1 \over 2\pi} \log |\gamma_p|^2$. For the conformally coupled scalar then $E_p = p$ and we have the entropy of a free gas  in $H^3$.
In other words, in the conformal case the entanglement Hamiltonian coincides with the Hamiltonian of
the field theory on $R \times H^3$ \refs{\CasiniKT,\Entsph}.

\subsec{Computing the Entropy} 

With the density matrix \finre\ we can calculate the entropy associated to each particular set of
$SO(1,3)$ quantum numbers
\eqn\entmode{
S(p,\nu) = - \Tr \rho_L(p) \log \rho_L(p)= -\log(1-|\gamma_p|^2)-{|\gamma_p|^2 \over 1-|\gamma_p|^2} \log |\gamma_p|^2
}
The final entropy is then computed by summing \entmode\ over all the states. This sum translates
into an integral over $p$ and a volume integral over the hyperboloid. In other words, we use
the density of states on the hyperboloid:
\eqn\entfinal{
S(\nu)=V_{H^3}\int dp {\cal D}(p) S(p, \nu)
}
The density of states for radial functions on the hyperboloid is known for any dimensions \BytsenkoBC. For example, for $H^3$, ${\cal D}(p)= { p^2 \over 2\pi^2 } $.
Here $V_{H^3}$ is the volume of the hyperboloid. This is of course infinite.
This infinity is arising because we are taking the entangling surface all the way to $\eta =0$.
We can regularize the volume with a large radial cutoff in $H^3$. This should roughly correspond to
putting the entangling surface at a finite time. Since we are only interested in the coefficient of the
logarithm, the precise way we do the cutoff at large volumes should not matter.
The volume of a unit size  $H^3$ for radius less that $r_c$ is given by
\eqn\volhyp{
V_{H^3}=V_{S^2} \int_0^{r_c} dr \sinh^2 r \sim 4\pi \left( {e^{2 r_c} \over 8}-{r_c\over 2}  \right)
}
The first term goes like the area of the entangling surface.
The second one involves the logarithm of this area.
 We can also identify
$r_c \to - \log \eta $. This can be understood more precisely as follows. If we fix a large  $t_L$ and we go to
large $r_L$, then we see from \angsfo \anaco\ that the corresponding surface would be at an $\eta \propto e^{ -r_L}$, for
large $r_L$.
Thus, we can confidently extract the coefficient $c_6$ in \entir . For such purposes we can define
$V_{H^3 {\rm reg} } = 2 \pi $. The leading area term , proportional to $e^{ 2 r_c}$ depends on the details
of the matching of this IR cutoff to the proper UV cutoff. These details can change its coefficient. 

\ifig\entrfree{
Plot of the entropy $S_{intr}/S_{intr,\nu=1/2}$ of the free scalar field, normalized to the conformally coupled scalar, versus its mass parameter squared. The minimally coupled massless case corresponds to $\nu^2=9/4$, the conformally coupled scalar to $\nu^2=1/4$ and for large mass (negative $\nu^2$) the entropy has a decaying exponential behavior.} {\epsfxsize5in\epsfbox{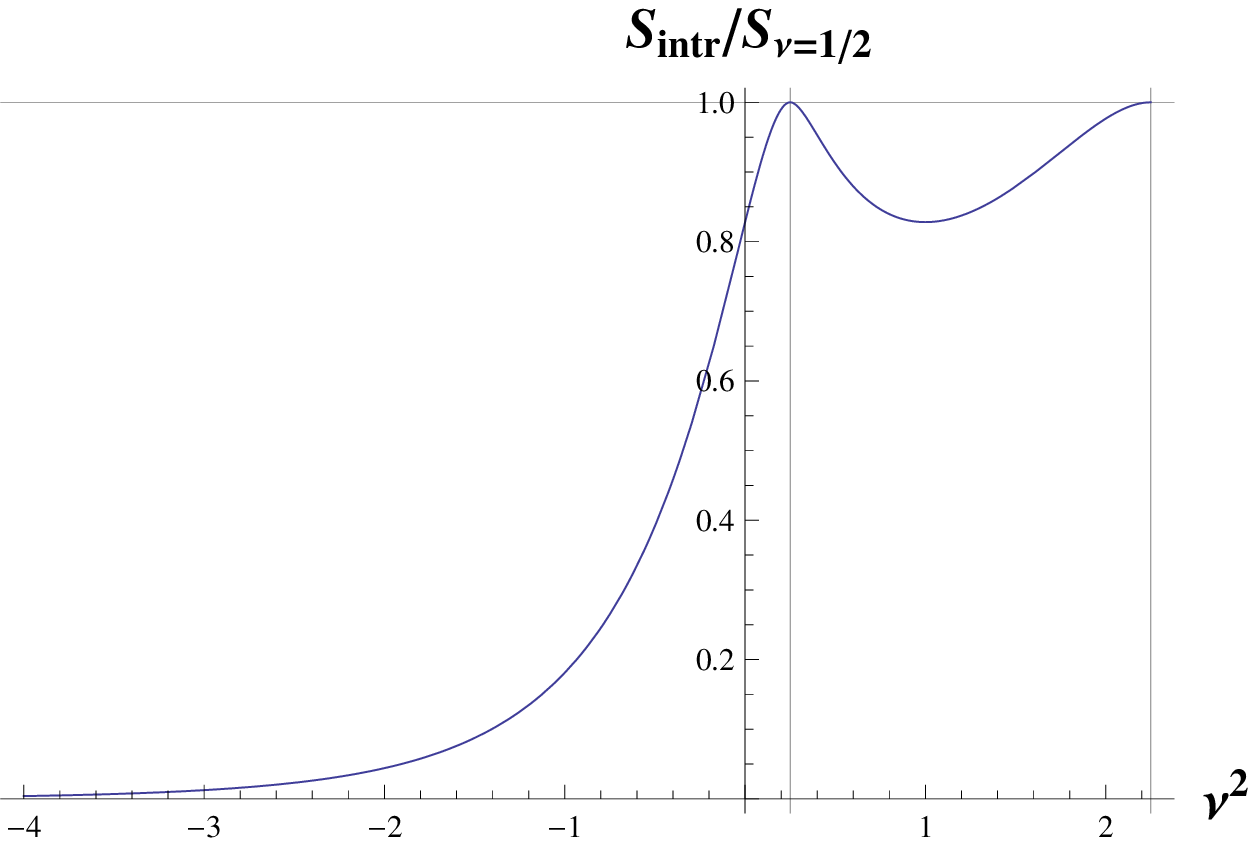}}

Thus, the final answer for the logarithmic term of the entanglement entropy is
\eqn\entelgo{ \eqalign{
S =& ~ c_6 \log \eta + {\rm other ~terms} \cr
S_{\rm intr } \equiv & ~c_6 = { 1 \over \pi } \int_0^\infty dp  \, p^2  S(p,\nu)
}}
with $S(p,\nu)$ given in \entmode , \eigen . This is plotted in \entrfree .

\subsec{ Extension to general dimensions }

These results can be easily extended to a real massive scalar field in any number of dimensions $D$. Again we
have hyperbolic $H^{D-1}$ slices and the decomposition of the time dependent part of the
wavefunctions is identical, provided that we replace $\nu$ by the corresponding expression in $D$
dimensions
\eqn\nud{
\nu^2 = { (D-1)^2 \over 4 } -  { m^2 \over H^2 }
}
Then the whole computation is identical and we get exactly the same function $S(p,\nu)$ for each mode.
The final result involves integrating with the right density of states for hyperboloids in $D-1$
 dimensions which is \BytsenkoBC
\eqn\denst{ \eqalign{
& {\cal D}_2(p)= {p \over 2\pi} \tanh \pi p,~~ {\cal D}_3(p)= {p^2 \over 2 \pi^2}\cr
& {\cal D}_{D-1}(p) = {p^2 + \left( {D-4 \over 2} \right)^2 \over 2 \pi (D-3)} {\cal D}_{D-3}(p) = {
 2 \over ( 4 \pi)^{ D-1 \over 2} \Gamma( { D -1 \over 2 } ) } { |\Gamma( i p + {D \over 2 } -1 ) |^2 \over
 |\Gamma(i p) |^2 } ,
 \cr
&  -{\bf L_{H^{D-1}}} Y_p = \left(p^2+ \left({D-2 \over 2} \right)^2\right)Y_p
}}
We also need to define the regularized volumes of hyperbolic space in $D-1$ dimensions. They
are related to the volume of spheres
\eqn\volregd{\eqalign{
V_{H^{D-1}, {\rm reg} } & = \cases{\displaystyle{(-1)^{D \over 2} \,  V_{S^{D-1}}\over \pi} & $D$ even \cr \displaystyle{(-1)^{D-1 \over 2} \, V_{S^{D-1}} \over 2}& $D$ odd}
 ~~, ~~~~~~~~~~~~~ V_{S^{D-1} }
 = { 2 \pi^{ D \over 2 } \over \Gamma( { D \over 2 } ) }
}}
When $D$ is even, we defined this regularized volume as minus the coefficient of $\log\eta$. When
$D$ is odd, we defined it to be the finite part after we extract the divergent terms. A derivation of these volume formulas is given in appendix B.
Then the final expression for any dimension is
\eqn\entanglr{
S_{\rm intr } = V_{H_{D-1}, {\rm reg} } \int_0^\infty dp {\cal D}_{D-1}(p) S(p, \nu)
}
with the expressions in \volregd , \denst , \entmode , \eigen , \nud .
We have defined  $S_{\rm reg}$   as
\eqn\defsre{\eqalign{
S   = &  S_{\rm intr} \log \eta + \cdots ~~~~~~~~~ {\rm for}~~ D~~{\rm even}
\cr
S = &  S_{\rm intr }  + \cdots  ~~~~~~~~~~~~~~~~ {\rm for}~~ D~~{\rm odd}
}}
where the dots denote terms that are UV divergent or that go like powers of $\eta$ for small $\eta$.

\subsec{R\'enyi Entropies}

We can also use the density matrix to compute the R\'enyi entropies, defined as:
\eqn\renyi{
S_q={1\over 1-q} \log \Tr \rho^q,~~q>0
}

We first calculate the R\'enyi entropy associated to each $SO(1,3)$ quantum number. It is given by:
\eqn\renpmod{
S_q(p,\nu)={q \over 1- q}\log(1-|\gamma_p|^2)-{1\over 1-q}\log(1-|\gamma_p|^{2q})
}
Then, just like we did for the entanglement entropy (which corresponds to $q\to1$), one integrates \renpmod\ with the density of states for $D-1$ hyperboloids:
\eqn\finreny{
S_{q,\rm intr } = V_{H^{D-1}, {\rm reg} } \int_0^\infty dp {\cal D}_{D-1}(p) S_q(p, \nu)
}
With $S_{q,\rm intr}$ being the finite term in the entropy, for odd dimensions, and the term that multiplies $\log \eta$, for even dimensions.

\subsec{Consistency checks: conformally coupled scalar and large mass limit}

As a consistency check of \entfinal, we analyze the cases of the conformally coupled scalar, and of masses much bigger than the Hubble scale.

{\bf Conformally coupled scalar}

For the conformally coupled scalar in any dimensions we need to
 set the mass parameter to $\nu=1/2$. The entropy should be the same as that of flat space. For a spherical entangling surface, the universal term is $g_e \log \epsilon_{UV}/R$ for even dimensions, and is a finite number, $g_o$, for odd dimensions \refs{\CasiniKT,\Entsph}. The only difference here is that we are following a surface of constant comoving area, so its radius is given by $R=R_c/(H\eta)$. So, one sees that the term that goes like $\log \eta$, in even dimensions, has the exact same origin as the $UV$ divergent one; in particular, we expect $c_6=g_e$ for the four dimensional case, and $g_o$ is the finite piece in the three dimensional case.

{\it Four dimensions:}

The entropy is given by \entelgo
\eqn\conffour{
S_{\rm intr} = { 1 \over \pi } \int_0^\infty dp {p^2\over 2\pi^2} S\left(p, {1 \over 2 } \right) = {1 \over 90}
}
This indeed coincides with the coefficient of the logarithm in the flat space result
 \CasiniKT .

{\it Three dimensions:}

The entropy is given by:
\eqn\confthr{ \eqalign{
S_{\rm intr} = &  V_{H^2,reg} \int { d^2 p \over (2 \pi )^2 } \tanh \pi p  S\left(p,{ 1 \over 2 } \right) = -
 \int_0^\infty p dp \tanh \pi p  S\left(p,{1 \over 2 } \right) =
\cr
=&
   {3 \zeta (3) \over 16 \pi ^2}-{\log (2) \over 8}
}}
This corresponds to half the value computed in \KlebanovUF , because there a complex scalar is considered, and also matches to half the value of the Barnes functions  in \Entsph.

{\it Conformally coupled scalar in other dimensions}

For even dimensions, $S_{\rm intr}$ has been reported for dimensions up to $d=14$ in \CasiniKT , and for odd dimensions, numerical values were reported up to $d=11$ \Entsph. Using \entanglr\ we checked that the entropies agree for all the results in \refs{\CasiniKT,\Entsph}.

{\bf  Large mass limit}

Here we show the behavior of the entanglement entropy for very large mass, in three and four dimensions. The eigenvalues of the density matrix
 as a function of the SO(1,3) Casimir are given in terms of  \eigen . For large mass, there are basically two regimes, $0<p < |\nu|$ and $p>|\nu|$
\eqn\eigenv{
|\gamma_p|^2= \cases{ e^{-2\pi |\nu|} & $0<p < |\nu|$\cr
e^{-2\pi p} & $ |\nu|< p $}
}
In this regime we can approximate $|\gamma|\ll 1 $ everywhere and the entropy per mode is
\eqn\enpm{
S(p) \sim  - |\gamma_p|^2 \log |\gamma_p|^2
}

Most of the contribution will come from the region $p< |\nu|$, up to $1/\nu$ corrections.
 This gives
\eqn\contrs{\eqalign{
&{S_{\rm intr}   \over V_{H^{D-1}, {\rm reg} } }  \sim  \int_0^\nu dp {\cal D}(p) S(p) \sim
 (2 \pi \nu e^{-2 \pi \nu}) \int_0^\nu dp {\cal D}(p)=\cases{{\nu^3 \over 2} e^{-2\pi\nu} & $d=3$ \cr {\nu^4 \over 3 \pi} e^{-2\pi\nu} & $d=4$}
 }}
which is accurate up to multiplicative factors of order $ (1 + {\cal O}(1/\nu ) )$.

\subsec{Low mass range: $1/2<\nu \le3/2$}

In this low mass range the expansion of the field involves an extra mode besides the ones we discussed so far \SasakiYT.
This is a mode with a special value of $p$. Namely $p = i (\nu - \half ) $. This mode is necessary because all the other modes, which have real $p$,  have
wavefunctions whose leading asymptotics vanish on the $S^2$ equator of the $S^3$ future boundary.
This mode has a different value for the Casimir (a different value of $p$) 
 than all other modes, so it cannot be entangled with them.
So we think  that this mode does not contribute to the  long range entanglement. It would be
nice to verify this more explicitly. 

Note that 
 we can analytically continue the answer we obtained for $\nu\leq 1/2$ to larger values. We obtain an answer which has
no obvious problems, so we suspect that this is the right answer for the entanglement entropy, even in this low mass range.
The full result is plotted in  \entrfree , and we find that for $\nu =3/2$, which is the massless scalar, we get exactly the same
result as for a conformally coupled scalar.

\newsec{Entanglement entropy from gravity duals. }

After studying free field theories in the previous section, we now
 consider strongly coupled field theories in de Sitter.
  We consider theories that have a gravity dual. Gauge gravity duality in de Sitter was 
  studied in 
  \refs{\HawkingDA,\BuchelTJ,\BuchelWF,\BuchelKJ,\AharonyCX,\BalasubramanianAM,\CaiMR,\RossCB,\AlishahihaMD,\BalasubramanianBG,\HirayamaJN,\BuchelEM,\HeJI,\HutasoitXY,\MarolfTG}, and references there in.
   When a field theory has a gravity dual, it was proposed in \RyuBV\
    that the entanglement entropy  is proportional
 to the area  of a minimal surface that ends on the entangling surface at the AdS boundary.
This   formula has passed many consistency checks. It is certainly valid in simple cases
   such    as  spherical entangling surfaces \CasiniKV.
Here we are considering a time dependent situation. It is then natural to use extremal surfaces but now in the full time
dependent geometry \HubenyXT . This extremality condition tells us how the surface moves in the time direction as it goes into the bulk.

First, we study a CFT in de Sitter. This is a trivial case since de Sitter is conformally flat, so we can
go to a conformal frame that is not time dependent and obtain the answer \refs{\RyuBV,\LiBT}. Nevertheless we will describe
 it in some detail because it is useful as a stepping stone for the non-conformal case.
We then consider non-conformal field theories in some generality.
We relegate to  appendix C the discussion of a special case corresponding to a non-conformal field theory
in  four dimensions that comes from compactifying a five dimensional
conformal field theory on a circle.

\subsec{Conformal field theories in de Sitter}

\ifig\adsfive{ The gravity dual of a CFT living on  $dS_4$. We slice $AdS_5$ with $dS_4$ slices.
 Inside the horizon we have an FRW universe with $H^4$ slices.
 The minimal surface is an $H^3$ that  lies on a constant global  time surface. The red line represents the
 radial direction of this $H^3$, and the $S^2$ shrinks smoothly at the tip.
} {\epsfxsize3in\epsfbox{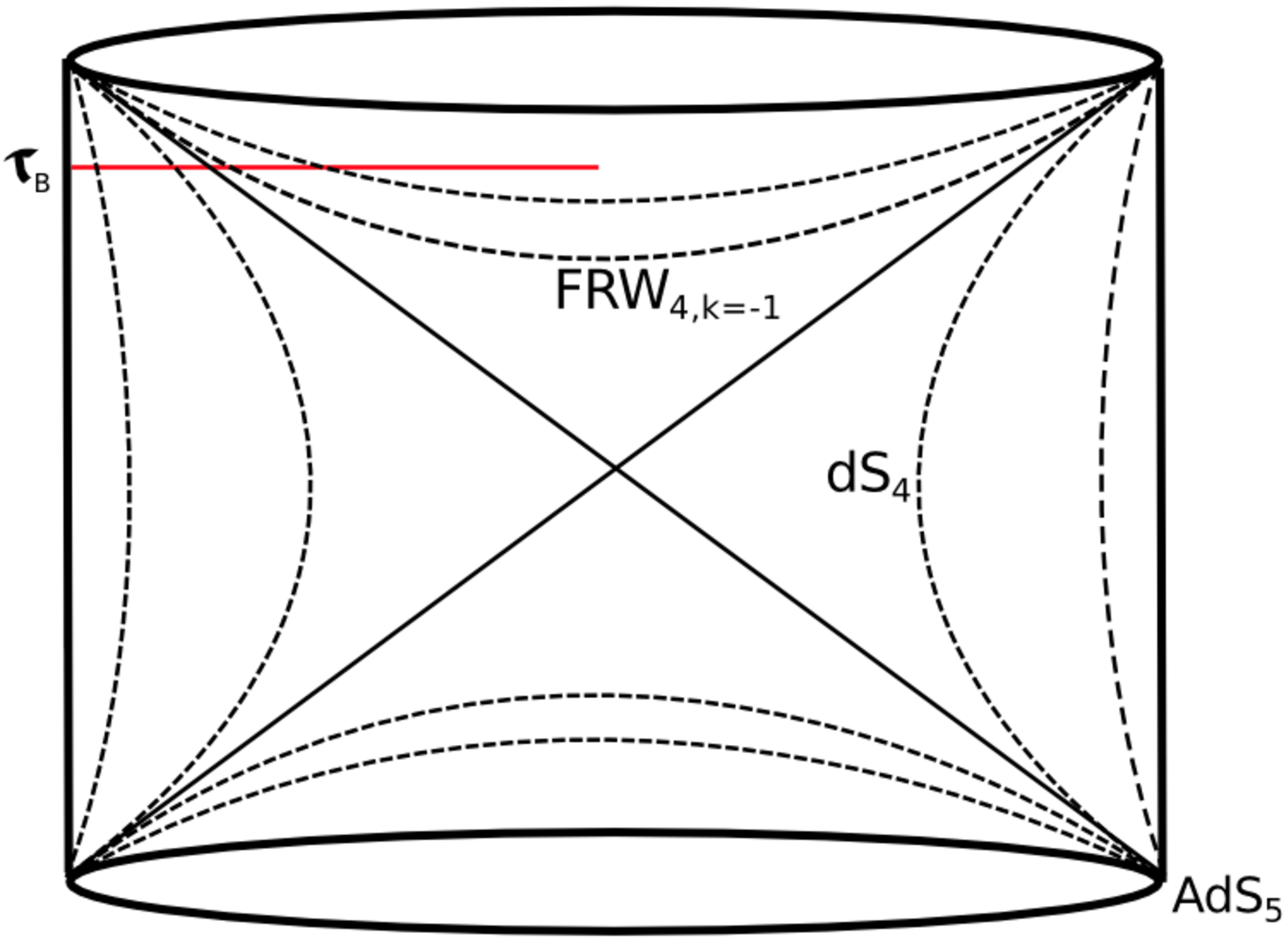}}

 As the field theory is defined in $dS_4$, it is convenient
  to choose a $dS_4$ slicing of $AdS_5$.
   These slices  cover only  part of the spacetime, see \adsfive .
     They cover the region outside the  lightcone of a point in the bulk. The interior region of this
     lightcone can be viewed as an FRW cosmology with hyperbolic spatial slices.

We then introduce the following coordinate systems:
\item{1.} Embedding coordinates
\eqn\emb{ \eqalign{
&-Y_{-1}^2-Y_0^2+Y_1^2+...+Y_4^2=-1 \cr
& ds^2=-dY_{-1}^2-dY_0^2+dY_1^2+...+dY_4^2}}
\item{2.} $dS_4$ and $FRW$ coordinates
 \item{2.1}  $dS$ slices
\eqn\dsads{\eqalign
{&Y_{-1}=\cosh \rho,~~~Y_0=\sinh \rho \sinh \tau,~~~Y_i=\sinh \rho \cosh \tau n_i \cr
&ds^2=d\rho^2+\sinh^2\rho(-d\tau^2+\cosh^2\tau(d\alpha^2+\cos^2\alpha d\Omega_2))
}}
\item{2.2}   $FRW$ slices. We substitute $\rho = i \sigma $ and $\tau = -i {\pi \over 2} + \chi$ in \dsads .
\eqn\frwads{\eqalign
{&Y_{-1}=\cos \sigma ,~~~Y_0= \sin \sigma \cosh \chi,~~~Y_i= \sin \sigma \sinh \chi n_i \cr
&ds^2=-d\sigma^2+\sin^2 \sigma (d\chi^2+\sinh^2\chi(d\alpha^2+\cos^2\alpha d\Omega_2))
}}
\item{3.} Global coordinates
\eqn\gloads{\eqalign
{&Y_{-1}=\cosh \rho_g \cos \tau_g,~~~Y_0=\cosh \rho_g \sin \tau_g,~~~Y_i=\sinh \rho_g n_i \cr
&ds^2=d\rho_g^2-\cosh^2\rho_g d\tau^2_g+\sinh^2\rho_g(d\alpha^2+\cos^2\alpha d\Omega_2)
}}

As the entangling surface we choose the $S^2$ at $\alpha=0$, at a large time $\tau_B$ and at $\rho=\infty$.
In terms of global coordinates the surface lies at a constant $\tau_{g}$, or at

\eqn\bdysurf{
{ Y_{0} \over Y_{-1}} = \sinh \tau_B = \tan \tau_{gB} , ~~~~~ Y_4=0
}
Its area is
\eqn\entcft{
A= 4\pi \int_0^{\rho_{gc}} \sinh^2\rho_g d\rho_g \sim 4 \pi \left( {e^{2\rho_{gc}} \over 8} - {\rho_{gc} \over 2} \right)
}
where $\rho_{gc}$ is the cutoff in the global coordinates. It is convenient to express this in terms
of the radial coordinate in the $dS$ slicing
 using $\sinh \rho_{g} = \sinh{\rho} \cosh \tau$.  In the large $\rho_{gc}$, $\rho_c$, $\tau_B$ limit
 we find  $\rho_{gc} \approx \rho_c + \tau_B   - \log 2 $.  Then \entcft\ becomes
\eqn\fencft{
A\sim 4\pi \left({e^{2\rho_c+2\tau_B}\over 16}  -{1\over 2} (\rho_c+\tau_B  ) \right)\sim 4\pi \left({1\over 16 (\eta~ \epsilon_{UV})^2}  +{1\over 2} (\log \epsilon_{UV} + \log \eta ~  ) \right)
}
We see that the coefficients of the two logarithmic terms are the same, as is expected in any CFT. Here
$\epsilon_{UV} = e^{ - \rho_c}$ is the cutoff in the de Sitter frame and $\eta \sim e^{ -\tau_B}$ is
de Sitter conformal
time.

\subsec{Non-conformal theories}

A simple way to get a non-conformal theory is to add a relevant perturbation to
a conformal field theory.
 Let us first discuss the possible Euclidean geometries. Thus we  consider theories on a sphere. In the interior we obtain a spherically symmetric metric and
profile for the scalar field of the form
\eqn\metrpro{
ds^2 = d\rho^2 + a^2(\rho) d \Omega_D^2 ~,~~~~~~~~~~\phi= \phi(\rho)
}
Some  examples were discussed in \refs{\BuchelIU,\BuchelWF,\HertogRZ} 
\foot{We are interpreting the solutions of \HertogRZ\ as
explained in appendix A of \MaldacenaUN .
 This geometry also appears in  decays of $AdS$ space \refs{\ColemanAW,\HertogRZ}.}.
If  the mass scale of the  relevant perturbation is small compared to the inverse
size of the sphere, the
dual geometry will be a small deformation of Euclidean $AdS_{D+1}$.
 Then we find that, at the origin, $a =  \rho + {\cal O}(\rho^3)$, and the sphere
shrinks smoothly. In this case we will say that we have the ``ungapped'' phase.
For very large $\rho$ we expect that $\log a \propto \rho $, if we have a CFT as the UV fixed point description.

\ifig\scfac{ The typical shape for the scale factor   for the gravity dual of a CFT perturbed by a
relevant operator in the ``ungapped'' phase. The region with negative $\rho^2$ corresponds to the FRW
region. In that region, we see that $\tilde a^2 = - a^2$ reaches a maximum value, $\tilde a_m$,
 and then contracts again
into a big crunch.
} {\epsfxsize3in\epsfbox{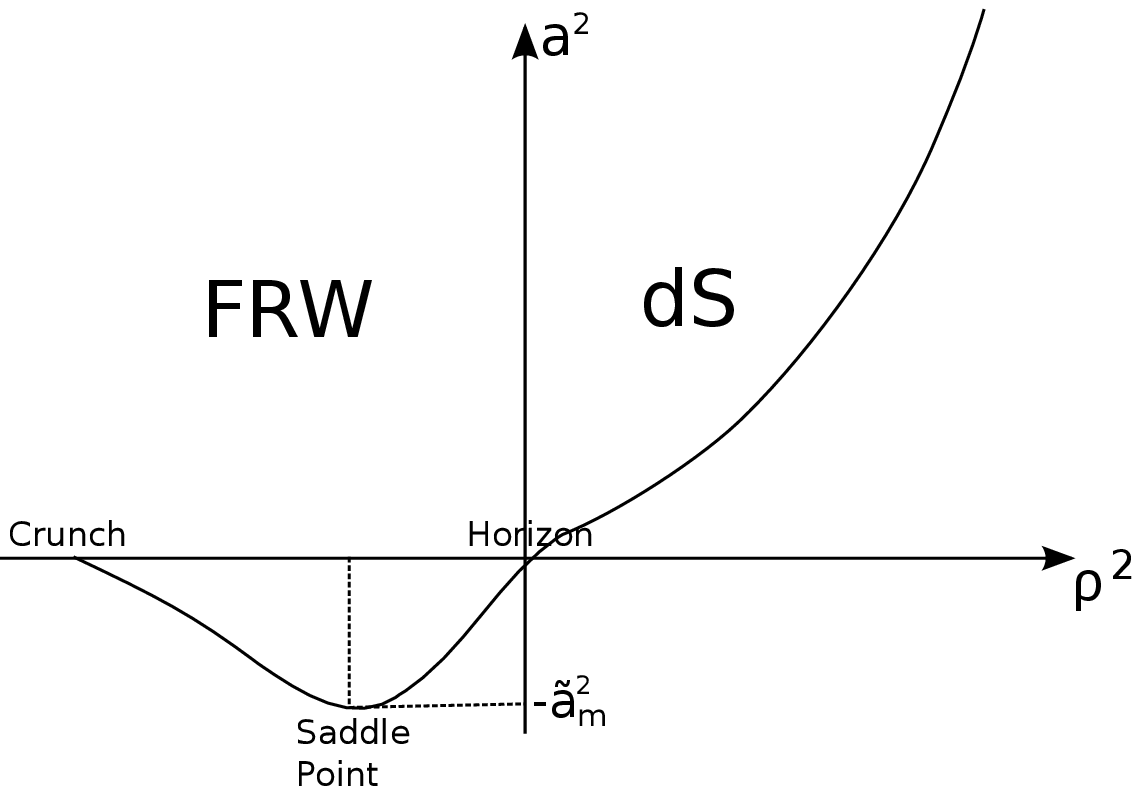}}

On the other hand, if the mass  scale of the relevant perturbation is large compared to the inverse size of
the sphere then the boundary sphere does not have to shrink when we go to the interior. For example,
the space can end before we get to $a=0$.
 This can happen in multiple ways. We could have an end of
the world brane at a non-zero  value of $a$.
Or some extra dimension could shrink to zero at this position. This typically happens for
the holographic duals of theories with a mass gap, especially if the mass gap is much bigger than $H$.
We call this the ``gapped'' phase.
See \refs{\AharonyCX, \CaiMR, \RossCB, \BalasubramanianBG, \MarolfTG, \BanadosDF, \BanadosGM} for some examples.
In principle,  the same field theory could display both phases as we vary the mass parameter
of the relevant perturbation. Then, there is  a large $N$ phase transition between the two regimes\foot{Since
we are at finite volume we might not have a true phase transition. In de Sitter, thermal effects will mix the
two phases. We will nevertheless restrict our attention to one of these phases at a time.}.

As we go to lorentzian signature, the ungapped case leads to a horizon, located at $\rho=0$.
 The metric is smooth if $a= \rho + {\cal O}(\rho^3)$. The region behind this horizon is obtained by
 setting $\rho = i \sigma$ in \metrpro\ and $d\Omega_D^2 \to - ds^2_{H_D}$.
This region looks like a Friedman-Robertson-Walker cosmology with hyperbolic spatial sections.

\eqn\cosmre{
ds^2 = - d\sigma^2 + (\tilde a(\sigma)  )^2 ds^2_{H_D} ~,~~~~~~~~~\tilde a(\sigma) \equiv -i a(i \rho) 
}
If the scalar
field is non-zero at $\rho=0$ we typically find that a singularity develops at a non-zero value of $\sigma$,
with the scale factor growing from zero at $\sigma =0$ and then decreasing again at the big crunch singularity. The scale factor then achieves a maximum somewhere in between, say at $\sigma_m$. See \scfac .

 We can choose
global coordinates for $dS_D$
\eqn\globc{
ds_{dS_D} = - d\tau^2 + \cosh^2 \tau ( \cos^2 \alpha d\Omega_{D-2} + d\alpha^2 )
}
We pick the entangling surface to be the $S^{D-2}$ at $\alpha =0$ and some late time $\tau_B$.
We assume that the surface stays at $\alpha =0$ as it goes into the bulk. In that case
we simply need to find  how $\tau$ varies as a function of $\rho$ as we go into the interior.
We need to minimize the following action
\eqn\actwom{
S = { V_{S^{D-2}}  \over 4 G_N} \int  ( a \cosh \tau )^{D-2} \sqrt{ d\rho^2 - a^2 d\tau^2 }
}

The equations of motion simplify if we assume $\tau$ is very large and we can approximate
$\cosh \tau \sim \half e^{ \tau }$. In that case the equations of motion give
a first order equation for $y \equiv  { d \tau \over d \rho }$.

\subsec{Non-conformal theories - gapped phase}

In the gapped phase, we can solve the equation for $y$. Inserting that back into the action will
give an answer that will go like $e^{ (D-2) \tau_B}$ times some function which depends on the details of the
solution. Thus, this produces just an area term. We can expand the action in powers of $e^{ - 2 \tau}$ and
obtain corrections to this answer. However, if the solution is such that the range of variation of $\tau$ is finite in the the interior, then
we do not expect that any of these corrections gives a logarithmic term (for even $D$) or a finite
term (for odd $D$).
Thus, in the gapped phase we get that
\eqn\reng{
S_{\rm intr } = 0
}
to leading order. The discussion is similar to the one in \refs{\KlebanovWS, \KlebanovYF} for a large entangling surface.

\subsec{Non conformal theories - ungapped phase }

In the ungapped phase, something more interesting occurs. The surface goes all the way to the horizon at
$\rho =0$. Up to that point the previous argument still applies and we expect no contributions to the
interesting piece of the entropy from the region $\rho > 0$.

When the surface goes into the FRW region note that the $S_{D-2}$ can shrink to zero at the origin
of the hyperbolic slices. If we call $\rho = i \sigma$ and $\tau = \chi - i \pi/2$, then we see that
the metric of the full space has the form
\eqn\metrf{
ds^2 = - d\sigma^2 + ( \tilde a(\sigma) )^2 [ d\chi^2 + \sinh^2 \chi ( \cos^2 \alpha d\Omega_{D-2} + d\alpha^2 )
]
}
where $\tilde a(\sigma ) = - i a(i \sigma )$ is the analytic continuation of $a(\rho)$.
We expect that the surface extends up to $\chi =0$ where the $S^{D-1}$ shrinks smoothly. Clearly this is
what was happening in the conformal case discussed in the previous subsection.  Thus, by continuity we expect that this also happens in
this case.

More explicitly, in this region we can write the action \actwom ,
\eqn\actnf{
S = { V_{S^{D-2}}  \over 4 G_N} \int  ( \tilde a \sinh \chi )^{D-2} \sqrt{ - \left( { d \sigma
\over d \chi } \right)^2 + \tilde a^2  }
}
\ifig\plottwo{The holographic setup for a non conformal field theory on de Sitter
in the ``ungapped'' phase.
We again have a region with  $dS_4$ slices and an FRW region with hyperbolic, $H_4$, slices.
The extremal  surface that computes the entanglement entropy
goes through  the horizon into the FRW region. There it approaches the
slice with maximum scale factor $a= a_m$.
} {\epsfxsize3.5in\epsfbox{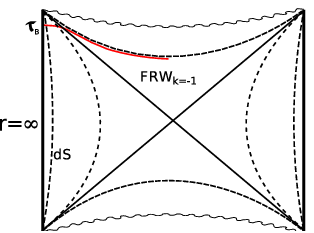}}
If we first set ${ d \sigma \over d \chi } =0$, we can extremize the area by sitting at $\sigma_m$ where
$\tilde a=\tilde a_m$, which is the maximum value for $\tilde a $.
We can then include small variations around this point. We find that we get exponentially increasing or
decreasing
solutions as we go away from $\sigma_m$. Since the solution needs to join into
 a solution with a very
large value of $\tau_B$, we expect that it will start with a value of $\sigma$ at $\chi=0$ which is
exponentially close to $\sigma_m$. Then the solution stays close to $\sigma_m$ up to $\chi \sim \tau_B$ and
then it moves away and approaches $\sigma \sim 0$. Namely, we expect that for $\sigma \sim 0$ the
solution will behave as
$\chi = \tau_B - \log \sigma + $rest, where the rest has an expansion in powers of $e^{ - 2 \tau_B}$.
This then joins with the solution of the form
$\tau = \tau_B - \log \rho +$rest in the $\rho>0$ region.
The part of the solution which we denote as ``rest'', has a simple expansion in powers of
$e^{ - 2 \tau_B}$, with
the leading term being independent of $\tau_B$. All those terms are not expected to contribute to the
interesting part of the entanglement entropy.
The qualitative form of the solution can be found in figure \plottwo .

The interesting part of the entanglement entropy comes from the region of the surface that sits near
$\sigma_m$. In this region the entropy behaves as
\eqn\acsby{
S = { \tilde a_m^{D-1}  \over 4 G_N } V_{H^{D-1}}  ~~\longrightarrow ~~~S_{\rm intr} =  { \tilde a_m^{D-1}  \over 4 G_N } V_{H^{D-1} {\rm reg }}
}
Here we got an answer which basically goes like the volume of the hyperbolic slice $H^{D-1}$. This
should be cutoff at some value $\chi \sim \tau_B$. We have extracted the log term or the finite term,
defined as the regularized volume.

Thus \acsby\ gives the final expression for the entanglement entropy computed using the gravity dual.
We see that the final expression is very simple. It depends only on the maximum value, $\tilde a_m$,
of the scale factor in the FRW region.

Using this holographic method, and finding the precise solution for the extremal surface 
 one can also
compute the coefficient $c_5$ in \entir\ (or analogous terms in general dimensions). But we will 
not do that here. 

In appendix C we discuss a particular example in more detail.
 The results agree with the general discussion we had here.
\newsec{Discussion}

In this paper we have computed the entanglement entropy of some quantum field theories in
de Sitter space. There are interesting features that are not present in the flat space case.
In flat space, a massive theory does not lead to any long range entanglement.
On the other hand, in de Sitter space particle creation
 gives rise to a long range contribution to the
entanglement. This contribution is specific to de Sitter space and does not have a flat
space counterpart. We   isolated this interesting part by considering a very large surface and
focusing on the terms that were either logarithmic (for even dimensions) or constant (for odd dimensions)
as we took the large area limit.

In the large area limit the computation can be done with relative ease thanks to a special SO(1,D-1) symmetry
that arises as we take the entangling surface to the boundary of $dS_D$. For a free field, this symmetry
allowed us to separate the field modes so that the entanglement involves only two harmonic oscillator
degrees of freedom at a time. So the density matrix factorizes into a product of density matrices for
each pair of harmonic oscillators. The final expression for the entanglement entropy for a free field
was given in \entanglr . We checked that it reproduces the known answer for the case of a conformally
coupled scalar. We also saw that in the large mass limit the entanglement goes as $e^{ - m/H}$ which
is due to the pair creation of massive particles. Since these pairs are rare, they do not 
produce much entanglement.

We have also studied the entanglement entropy in theories that have gravity duals. The interesting
contribution to the entropy only arises when the bulk dual has a horizon. Behind the horizon there
 is an FRW region with hyperbolic cross sections. The scale factor of these hyperbolic cross sections
 grows, has a maximum, 
  and then decreases again. The entanglement entropy comes from a surface that
 sits within  the hyperbolic slice at the time of  maximum expansion. This gives a simple formula
 for the holographic entanglement entropy \acsby .
 From the field theory point of view, it is an $N^2$ term. Thus, it comes from the long range
 entanglement of colored fields. It is particularly interesting that the long range entanglement comes from the FRW cosmological
 region behind the horizon. 
 This suggests that this FRW cosmology is indeed somehow contained in the
 field theory on de Sitter space \refs{\MaldacenaUN,\BarbonTA} . 
  More precisely, it is contained in colored modes that are correlated
 over superhorizon distances. 
 
 In the gapped phase the order $N^2$ contribution to the long range entanglement
 entropy vanishes. We expect to have an order one contribution that comes from bulk excitations which
 can be viewed as color singlet massive excitations in the boundary theory. From such contributions we
 expect an order one answer which is qualitatively similar to what we found for free massive scalar fields above.

{\bf Acknowledgements }

We thank H. Liu, I. Klebanov, R. Myers, B. Safdi and A. Strominger for discussions.
J.M. was  supported in part by U.S.~Department of Energy
grant \#DE-FG02-90ER40542. G.L.P. was supported by
the Department of State through a Fulbright Science and Technology Fellowship and through the
U.S.~NSF under Grant No. PHY-0756966.

\appendix{A}{Bogoliubov coefficients}

Here we give the explicit form of the coefficients in \basis .
\eqn\coefba{ \eqalign{
\alpha^\sigma_R = &  {e^{\pi p} - i \sigma e^{-i \pi \nu} \over \Gamma(\nu+ i p +1/2)} ~,~~~~~~~~~~
\alpha^\sigma_L = \sigma {e^{\pi p} - i \sigma e^{-i \pi \nu} \over \Gamma(\nu+ i p +1/2)}
\cr
\beta^\sigma_R = & -  {e^{-\pi p} - i \sigma e^{-i \pi \nu} \over \Gamma(\nu- i p +1/2)}~,~~~~~~~~~~
\beta^\sigma_L = -  \sigma {e^{-\pi p} - i \sigma e^{-i \pi \nu} \over \Gamma(\nu- i p +1/2)}
}}
We also find
\eqn\formga{ \eqalign{
\gamma_{j \sigma } = & { \Gamma( \nu + i p + \half ) i  e^{\pi p + i \pi \nu } \over 4 \sinh \pi p } \pmatrix{
{ 1 \over  i e^{ \pi p + i \pi \nu } + 1 } & { 1 \over  i e^{ \pi p + i \pi \nu } - 1 }
\cr
{ 1 \over  i e^{ \pi p + i \pi \nu } + 1 } & - { 1 \over  i e^{ \pi p + i \pi \nu } - 1 } }_{j\sigma}
\cr
\bar \delta_{j \sigma } = & { \Gamma( \nu - i p + \half ) i  e^{\pi p + i \pi \nu } \over 4 \sinh \pi p }
\pmatrix{
{ 1 \over  i e^{ \pi p + i \pi \nu } + e^{ 2 \pi p }  } & { 1 \over  i e^{ \pi p + i \pi \nu } -  e^{ 2 \pi p }  }
\cr
{ 1 \over  i e^{ \pi p + i \pi \nu } + e^{ 2 \pi p }  } & - { 1 \over  i e^{ \pi p + i \pi \nu } -  e^{ 2 \pi p }  } }_{j\sigma}
}}
these were used to obtain \solex .

We define $c_R$ and $c_L$ via \bogol\ and the state in \finalc . We demand that
 $c_R | \Psi \rangle = \gamma c^\dagger_L | \Psi \rangle, c_L | \Psi \rangle = \gamma c^\dagger_R | \Psi \rangle$.
Using \bogol\ and
denoting $m_{RR}=m_{LL}=\rho$, $m_{RL}= \zeta $ these two conditions become
\eqn\conds{\eqalign{
&(u \rho + v - \gamma v \zeta ) b^\dagger_R+(u \zeta -\gamma v \rho - \gamma u)b^\dagger_L=0 \cr
&(\bar u \zeta - \gamma \bar u -\gamma \bar v \rho) b^\dagger_R +(\bar u \rho + \bar v - \gamma \bar v \zeta) b^\dagger_L=0
}}
which imply that each of the coefficients is zero.

From the structure of \conds , one sees that under the substitution $u \to \bar u$, $v \to \bar v$ we have the same set of equations. If one tries to solve them together then ${u \over v} = {\bar u \over \bar v}$; hence this ratio must be real. One can show that this is indeed the case and $\gamma$  is given by \eigen .

\appendix{B}{Regularized volume of the Hyperboloid}

Here we calculate the regularized volume of a hyperboloid in $D-1$ dimensions. We have to consider the cases of $D$ even and $D$ odd separately. First, note that the volume is given by the integral:
\eqn\volhyp{
V_{H^{D-1}}=V_{S^{D-2}}\int_0^{\rho_c} d\rho (\sinh \rho)^{D-2}
}

Now we expand the integrand:
\eqn\volexp{
{V_{H^{D-1}}\over V_{S^{D-2}}}={1 \over 2^{D-2}}\int_0^{\rho_c} d\rho \sum_{n=0}^{D-2} {D-2 \choose n}(-1)^n e^{(D-2-2n)\rho}
}

But the integral of any exponential is given by:
\eqn\expint{
\int_0^{\rho_c} d\rho ~e^{a \rho}=-{1\over a} +\cases{0,& $a<0$\cr {\rm divergent}, &$a>0$}
}

Now we treat even or odd dimensions separately.

{\bf Even $D$:} Here, the integrand of \volexp\ contains a 
 term independent of $\rho$ in the summation, which gives rise to 
  the logarithm (a term linear in $\rho_c$). 
  The term we are interested in corresponds to setting $n=D/2 -1$:
\eqn\logter{
V_{H^{D-1},{\rm reg}}={(-1)^{D\over 2}\over {D-2 \over 2}}{D-2 \choose {D-2\over 2}}V_{S^{D-2}}={(-1)^{D\over 2}\pi^{D-2\over 2}\over {D-2 \over 2}}{(D-2)! \over \left({D-2\over 2}\right)!^3}={(-1)^{D\over 2} \over \pi}V_{S^{D-1}}
}

{\bf Odd $D$:} Now, there is no constant term in the integrand of \volexp . 
Performing the summation in \volexp , and using \expint , we get:
\eqn\logter{
V_{H^{D-1},{\rm reg}}=-{1\over 2^{D-2}}\sum_{n=0}^{D-2} {D-2 \choose n}{(-1)^n\over D-2-2n}V_{S^{D-2}}=\pi^{D-2\over 2}\Gamma\left[-{D-2\over 2}\right]={(-1)^{D-1\over 2} \over 2}V_{S^{D-1}}
}

A more direct way to relate the regularized volumes of hyperbolic space to volume of the
corresponding spheres is by a shift of the integration contour. We change $\rho_c \to \rho_c + 
i \pi $. This does not change the constant term, but we get an $i \pi$ from the log term. 
We then shift the contour to run from $\rho=0$ along $\rho = i \theta$, with $0 \leq \theta \leq 
\pi $ and then from $i\pi $ to $i\pi + \rho_c$. The $\theta$ integral gives the volume of 
a sphere and the new integral with $Im(\rho) = \pi$ gives an answer which is either the same
or minus the original integral. 
The fact that these regularized volumes are given by volume of spheres is related to the 
analytic continuation between $AdS$ and $dS$ wavefunctions \refs{\MaldacenaVR,\HarlowKE}.

\appendix{C}{Entanglement entropy for conformal field theories on $dS_4 \times S^1$}

Let us first discuss the gravity dual in the Euclidean case. The boundary is $S^4 \times S^1$.
This boundary also arises when we consider a thermal configuration for the field theory on $S^4$.
We will consider antiperiodic boundary conditions for the fermions along the $S^1$. There
are two solutions. One is AdS with time compactified on a circle. The other is the Schwarzschild AdS
black hole. Depending on the size of the circle one or the other solution is favored \refs{\HawkingDH,\WittenZW}. Here we want to continue $S^4 \to dS^4$.  An incomplete list of
references where these geometries were explored is \refs{\AharonyCX, \CaiMR, \RossCB, \BalasubramanianBG, \MarolfTG, \BanadosDF, \BanadosGM} .  

As a theory on $dS^4$ we have
a scale set by the radius of the extra spatial circle. At large $N$ we have a sharp phase transition. At finite
$N$ we can have tunneling back and forth between these phases. Here we restrict attention to
one of the phases, ignoring this tunneling.
The Schwarzschild AdS solution   looks
 basically like the gapped solutions we discussed in general above.
Here, the $S^4$, or the $dS^4$, never shrinks to zero. It can be viewed as a bubble of nothing.
 On the other hand, the periodically identified $AdS_6$
solution gives the ungapped case, with the $S^4$ or $dS^4$ shrinking, which leads to a hyperbolic FRW
region behind the horizon.

{\it Gapped phase - Cigar geometry}

\ifig\schwads{ The gravity dual of a 5D CFT on $dS_4 \times S^1$. 
 The spacetime ends   at $r=r_h$, where the circle shrinks in a smooth fashion. 
  We display an extremal surface going from $\tau_B$ to the interior. 
} {\epsfxsize3in\epsfbox{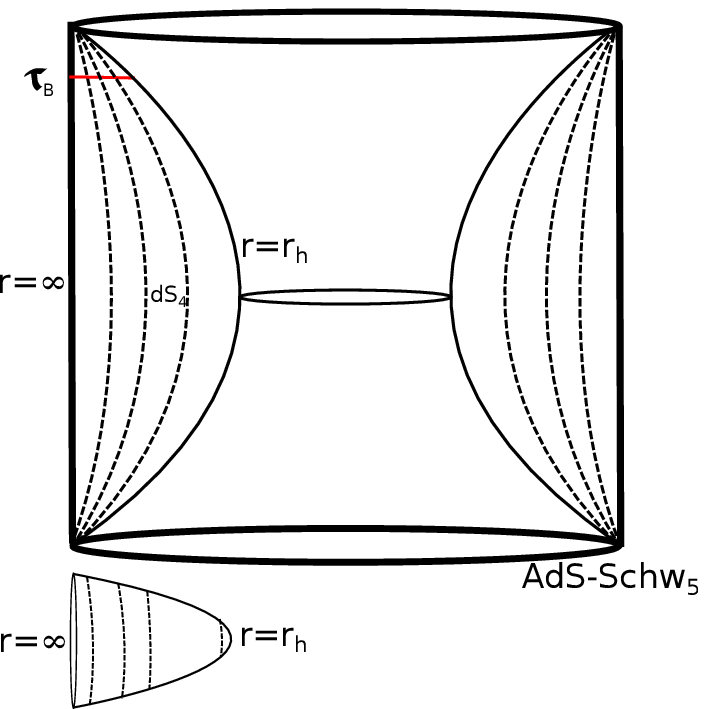}}

We now consider the cigar geometry. In the UV we expect to see the divergence structure to be that of a 5D CFT, but in the IR it should behave like a gapped 4D non-conformal theory.
The   metric is given by
\eqn\cigar{
ds^2=f d\phi^2+r^2 ds^2_{dS^4}  +{ dr^2 \over f},~~~~~~f=1+r^2-{m\over r^3}
}
The period $\beta$ of the $\phi$ circle is given in terms of $r_h$, the largest root of $f(r_h)=0$, by
\eqn\coef{
\beta=
{4 \pi \over f'(r_h)}={4 \pi r_h \over2  +4 r_h^2}
}
Note that $\beta_{max}=\pi/\sqrt{2}$. This solution only exists for  $\beta \leq  \beta_{max}$.
This geometry is shown in \schwads .
 We consider an entangling surface which is an $S^2$ at a large value of $\tau_B$.

We need to consider the action
\eqn\func{
A \propto \int dr r^2 \cosh^2 \tau  \, \sqrt{1-f r^2 \tau'^2}
}
This problem was also discussed in \HubenyXT . 
Since we are interested in large $\tau_B$ we can approximate this by
\eqn\acapp{
A_{approx} = \int dr  r^2 e^{ 2 \tau } \sqrt{ 1  - f  r^2 (\tau')^2 }
}
 If the large tau approximation is valid throughout the solution then we see that the dependence on $\tau_B$
drops out from the equation and it only appears normalizing the action. In that case the full result is proportional
to the area, $e^{ 2 \tau_B}$, with no logarithmic term.
In the approximation \acapp, the equation of motion involves only $\tau'$ and $\tau''$.
So we can define a new variable $y \equiv \tau'$ 
and the equation becomes first order.  One can expand the equation for $y$ and get that $y$ has an expansion of the form
$y = [-2/(3r^3) + 10/(27r^5) - 4m/(14r^6) +\cdots]+ a (1/r^6 + \cdots ) $ where $a$ is an arbitrary coefficient representing
the fact that we have one integration constant.

This undetermined coefficient should be set by requiring that the solution is
smooth at $r= r_h$.  If one expands the equation around $r=r_h$, assuming the solution has a power series
expansion around $r_h$, then we get that $y$ should have a certain fixed value at $r_h$ and then all its powers
are fixed around that point. Notice that if $y = y_h + y'_h (r-r_h) + \cdots$, implies that
$\tau$ is regular around that point, since $(r-r_h) \propto 
 x^2 $ where $x$ is the proper distance from the tip.

The full solution can be written as
\eqn\solf{
\tau = \tau_B - \int_{r_h}^\infty y(r) dr
}
where $y$ is independent of $\tau_B$.

\ifig\plotone{ The regulated area $A_{\rm reg}$ is defined by $A_{\rm total}=e^{2\tau_B} A_{\rm reg} + A_{\rm div}$.
} {\epsfxsize3in\epsfbox{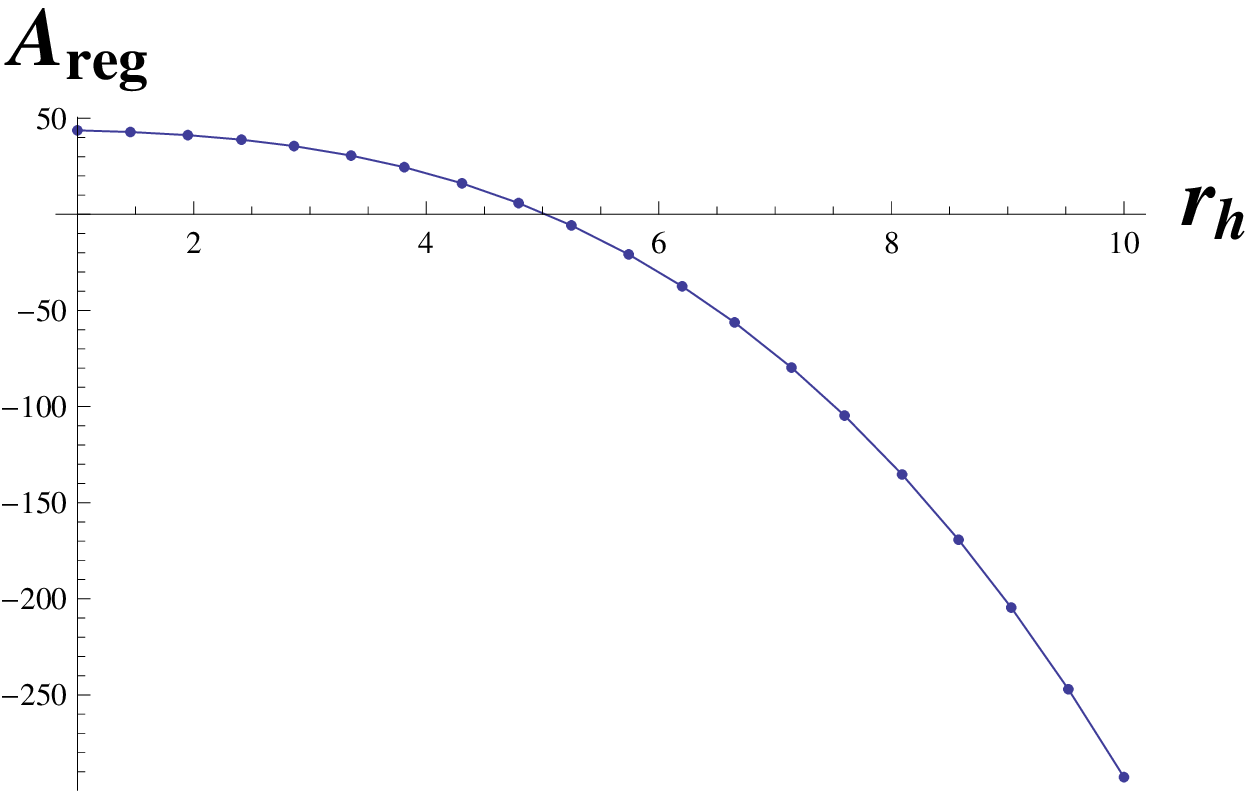}}
At  large $r$  we get $
\tau - \tau_B  = { 1 \over 3 r^2 } + {\cal O}(1/r^4)
$
and the action \acapp\  evaluates to
\eqn\evalca{
A_{approx} \propto   { e^{ 2 \tau_B } } \int dr  \left[ r^2 + { 4 \over 9 }  + {\cal O}\left({1\over r^2}\right) \right] \sim
  { e^{ 2 \tau_B } } \left( { r_c^3 \over 3 }  + { 4 \over 9 } r_c  + {\rm finite} \right)=e^{2\tau_B}A_{reg}+A_{div}
  }
  We see that we get the kind of UV divergencies we expect in a five dimensional theory, as expected.
 
These can be subtracted and we can compute the finite terms. These are plotted in \plotone\ as a function
of $r_h$.

So far, we have computed the finite term that grows like the area. By expanding \func\ to the next order
in the $e^{ - 2 \tau}$ expansion we can get the next term. The next term will give a constant value,
independent of the area. In particular, it will not produce a logarithmic contribution. In other words,
there will not be a contribution proportional to $\tau_B$.

In conclusion, in this phase, there is no logarithmic contribution to the entanglement entropy, at order
$1/G_N$ or $N^2$.

{\it Ungapped  phase - Crunching geometry}

Now the geometry is simply  $AdS_6$ with an identification.
 This construction is described in detail in \refs{\BanadosDF,\BanadosGM}.
  The resulting geometry has a big crunch singularity where the radius of the spatial circle shrinks to zero.
  This geometry is sometimes called `` topological black hole'', as a higher dimensional generalization of the BTZ solution in $3D$ gravity.

It is more convenient to use a similar coordinate system as the one used to describe the cigar geometry in the previous case. The metric is given by \cigar , with $m=0$. Those coordinates only cover the region outside of the lightcone at the origin,   $r= 0$. To continue into the $FRW$ region, one needs to use $r = i \sigma $ and $\tau=-i \pi/2 + \chi$ in \cigar .

The equation in the $r>0$ region is such that we can make the large $\tau $ approximation and
it thus reduces to a first order equation for $y = \tau' = { d \tau \over dr }$.
For small $r$, an analysis of the differential equation tells us that
\eqn\diffeq{
y \sim - { 1 \over  r } -2 r^3 + \cdots  +  b \left(r + {10-7b \over 2}r^3 + \cdots \right)
}
where there is only one undetermined coefficient (or integration constant) which is $b$ (it is really non-linear in $b$). This leads to
a $\tau$ which is
\eqn\forta{
\tau \sim - \log r - r^4 + \cdots  + c + b \left( r^2 + {10-7b \over 8}r^4 +\cdots \right)
}
where $c$ is a new integration constant. We expect that the evolution from this near horizon region to infinity
only gives a constant shift. In other words, we expect that $c = \tau_B + $constant.
 This constant
appears to depend on the value of $b$ that is yet to be determined.
We find that $b$ should
 be positive in order to get a solution that goes to infinity and is non-singular.

We are now supposed to analytically continue into the $FRW$ region.
For that purpose we set
$r = i \sigma $ and $\tau = - i \pi/2 + \chi $.
Thus the equation \forta\ goes into
\eqn\nefu{
\chi  \sim - \log \sigma - \sigma^4 + \cdots + c+ b \left( - \sigma^2 + {10-7b \over 8} \sigma^4 +\cdots  \right)
}
Then we are supposed to evolve the equation. It is convenient to change variables and write the Lagrangian in terms of $\sigma(\chi)$ as:
 \eqn\lagoth{
 A   \sim \int d \chi \sigma^2  \sinh^2 \chi  \sqrt{ - (\sigma')^2 +  \sigma^2 (1-\sigma^2) }
 }

 In this case,  at $\chi =0$ we can set any starting point value for $\sigma(\chi=0)$ and we have to
 impose that $\sigma'=0$. Then we get only one integration constant which is $\sigma( 0)$, as the second derivative
 is fixed by regularity of the solution to be $\sigma''(0) =- \sigma(0)( 3 -  4 \sigma(0)^2 ) $. We see that its sign depends on the starting value of $\sigma(0)$.

Since we want our critical surface to have a ``large'' constant value
 when we get to $\sigma\to 0$ as $\chi \to \infty$, we  need to tune the value of $\sigma(0)$ so that it gives
 rise to this large constant. This can be obtained by
  tuning the coefficient in front of $\sigma(0)$. This critical value of $\sigma(0)$ is easy to understand.
  It is a solution of the equations of motion with $\sigma'(\rho ) =0$ (for a constant $\sigma$), it is  a saddle
  point for the solution, located at $\sigma = \sqrt{3}/2$. If $\sigma$ is slightly bigger than the critical value, the minimal surface will collapse into the singularity, so we tune this value to be slightly less than the critical point.

 So, in conclusion, we see that the surface stays for a while at $\sigma \approx \sqrt{3}/2$ which is the critical point stated above.
 The value of the action \lagoth\ in this region is then 
 \eqn\valacr{
 { 3 \sqrt{3} \over 16 } \int_0^{\chi_c}  d\chi \sinh^2 \chi  =  { 3 \sqrt{3} \over 16 } \left[
 { e^{ 2 \chi_c } \over 8 } - { \chi_c \over 2 } + \cdots \right]
 }
 Here $\chi_c \sim \tau_B \sim \log \eta$ is the value of $\chi$ at the transition region.
 Thus, we find that the interesting contribution to the entanglement entropy is coming from the FRW region.

The transition region and the solution all the way to the $AdS$ boundary is expected to be universal and
its action is not  expected to contribute further logarithmic terms.

In conclusion, the logarithmic term gives
\eqn\entlo{
S_{\rm intr} = { R_{AdS_6}^4 \over 4 G_N } \beta { 4 \pi  3 \sqrt{3} \over 32}
}

Here we repeat this computation in another coordinate system which is non-singular at the 
horizon. 
We use Kruskal-like coordinates \refs{\BanadosDF,\BanadosGM,\CaiMR} . It also makes the numerical analysis much simpler. In terms of embedding coordinates for $AdS_6$:
\eqn\embsix{ \eqalign{
&-Y_{-1}^2-Y_0^2+Y_1^2+...+Y_5^2=-1 \cr
& ds^2=-dY_{-1}^2-dY_0^2+dY_1^2+...+dY_5^2}}
The Kruskal coordinates are given by:
\eqn\kruskal{
\eqalign
{&Y_{-1}={1+y^2 \over 1- y^2}\cosh \phi,~~~Y_5={1+y^2 \over 1- y^2}\sinh \phi,~~~Y_{0,...,4}={2 y_{0,...4} \over 1-y^2},~~~y^2\equiv -y_0^2+y_1^2+...+y_4^2\cr
&~~~~~~~~~~~~~~~~~~~~~~~~~ds^2={4 \over (1-y^2)^2}(-dy_0^2+...+dy_4^2)+\left({1+ y^2 \over 1-y^2}\right)^2 d\phi^2
}
}
In these coordinates, the $dS$ region corresponds to $0<y^2<1$ and the $FRW$ region to $-1<y^2<0$, with the singularity located at $y^2=-1$. The $AdS$ boundary is at $y^2=1$. 
We can relate the pair $(r, \tau)$ and $(\chi, \sigma)$, connected by the analytic continuation $(r=i\sigma ,~\tau=-i\pi/2+\chi)$ to $(y^2, y_0)$ by the formulas:
\eqn\conv{\eqalign{
&r={2\sqrt{y^2}\over 1- y^2},~~~\sinh \tau = {y_0\over\sqrt{y^2}}\cr
&\sigma ={2\sqrt{-y^2}\over 1- y^2},~~\cosh \chi = {y_0\over\sqrt{-y^2}}
}
}

The area functional gets simplified to:
\eqn\arkrus{
A=\int\sqrt{{16 (1+y^2)^2\over(1-y^2)^8}(y_0^2+y^2)\left[(d(y^2))^2+4 y_0 dy_0 d(y^2)-4y^2 (dy_0)^2\right]}
}
\ifig\plotthree{
We plot here the value of $y^2$ versus $y_0$. 
For small $y_0$ the solution starts very close to the surface of maximum expansion at $y^2 =-1/3$, stays there for a while and then they go into the AdS boundary at $y^2 =1$. 
The closer $y_0$ is to the saddle point $\tilde y_m^2=-1/3$, the longer the solution will stay 
on this slice,  giving a contribution that goes like the volume of an $H^3$. Then, at a time $y_0$ of the order of the time the surface reaches the boundary, it exits the $FRW$ region. The 
interesting (logarithmic) contribution to the entropy is coming  from the volume of the $H^3$ surface along the $FRW$ slice at $y^2 =-1/3$.
} {\epsfxsize3in\epsfbox{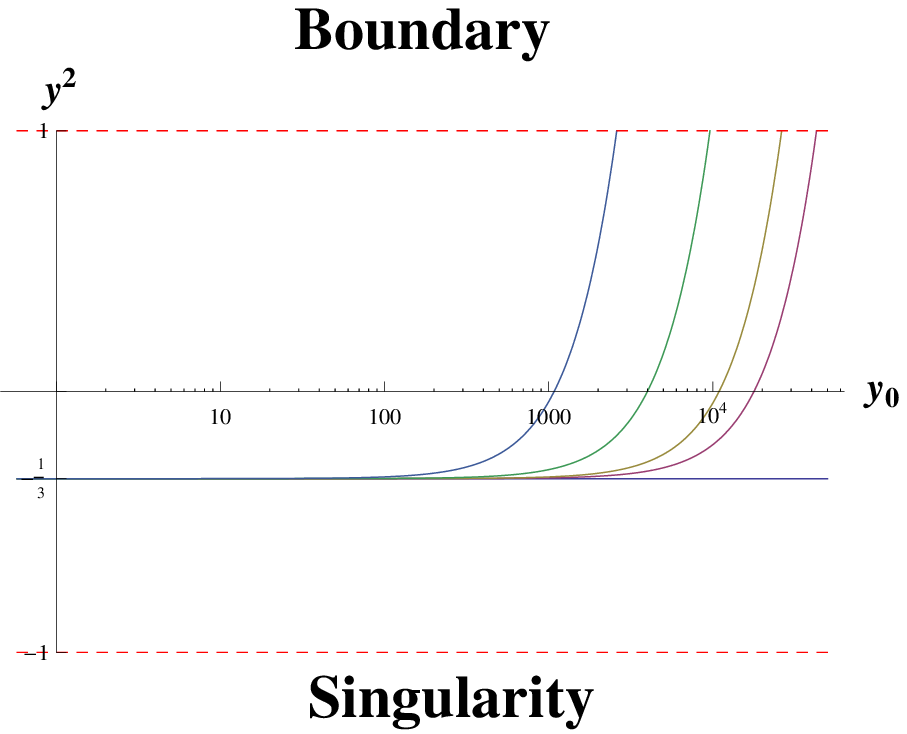}}

If one looks for the saddle point described in the $FRW$ coordinates, then one obtains $y^2=-1/3$. The same situation can be described in simple fashion in terms of these coordinates.
So $y^2\sim -1/3 $ for a large range of $y_0$, and it crosses to the $dS$-sliced region at a time of order $\tau_B$, in $dS$ coordinates, so part of the surface just gives the volume of an $H^3$, as in \valacr:
\eqn\arkr{
 A\sim {9\over 16}\int_0^{\cosh\chi_c\over \sqrt{3}}\sqrt{3 y_0^2-1}~ dy_0=  { 3 \sqrt{3} \over 16 } \left[
 { e^{ 2 \chi_c } \over 8 } - { \chi_c \over 2 } + \cdots \right]
}
Some plots for the minimal surfaces are shown in \plotthree . 

\listrefs
\bye